\documentclass[11pt,a4paper,epsfig]{article}

\usepackage{amsfonts}
\usepackage{epsfig}

\title{\textbf{ One-dimensional solitary waves in singular
deformations of $SO(2)$ invariant two-component scalar field theory
models}}
\author{A. Alonso Izquierdo$^{(a)}$ and J. Mateos Guilarte$^{(b)}$
\\ {\normalsize {\it $^{(a)}$ Departamento de Matematica
Aplicada}, {\it Universidad de Salamanca, SPAIN}}\\{\normalsize
{\it $^{(b)}$ Departamento de Fisica Fundamental and IUFFyM}, {\it
Universidad de Salamanca, SPAIN}}}

\date{}

\begin{document}

\maketitle

\begin{abstract}
In this paper we study the structure of the manifold of solitary
waves in some deformations of $SO(2)$ symmetric two-component
scalar field theoretical models in two-dimensional Minkowski
space. The deformation is chosen in order to make the analogous
mechanical system Hamilton-Jacobi separable in polar coordinates
and displays a singularity at the origin of the internal plane.
The existence of the singularity confers interesting and
intriguing properties to the solitary waves or kink solutions.
\end{abstract}

\section{Introduction}

Solitary waves are at the heart of astonishing phenomena in
diverse physical systems that can be modeled by non-linear
equations. For instance, they describe the behavior of interfaces
in magnetic materials \cite{Eschen} and in ferroelectric crystals
\cite{Jona} in Condensed Matter. Solitary waves have biotechnical
and biomedical applications \cite{Harris} and also seed the
formation of structures in Cosmology \cite{Vilenkin}.  In several
disguises, kinks, topological defects, domain walls, membranes,
solitary waves arise in many branches of Theoretical Physics
\cite{Koba,Olive}. For this reason, the search for solitary waves
in some types of PDE is an active topic in non-linear science.
Among these equations we find the non-linear Klein-Gordon
equation, extensively cited in the physical literature
\cite{Drazin,Rajaraman}. This equation, generalized to $N$-real
scalar fields, reads as follows:
\begin{equation}
\frac{\partial^2 \phi_a}{\partial t^2}-\frac{\partial^2
\phi_a}{\partial x^2} +\frac{\partial U}{\partial \phi_a}=0
\hspace{1cm} a=1,2, \dots , N \qquad , \label{eq:nlkg}
\end{equation}
where $U(\phi_1,\dots,\phi_N)$ is a potential function of the scalar
fields $\phi_a$, whereas $\frac{\partial U}{\partial \phi_a}$ are
non-linear functions of the fields. PDE (\ref{eq:nlkg}) can be
understood as the Euler-Lagrange equations associated with the
functional action
\[
S=\int d^2 x \left[\frac{1}{2}\sum_{a=1}^N \partial_\mu \phi^a
\partial^\mu \phi^a - U(\phi_1,...,\phi_N) \right]
\]
governing the dynamics of a (1+1) dimensional scalar field theory.
In this framework, \textit{a solitary wave is a localized
non-singular solution of the non-linear field equation
(\ref{eq:nlkg}) whose energy density, as well as being localized,
has space-time dependence of the form:
$\varepsilon(t,x)=\varepsilon(x-\mbox{v} t)$, where v is some
velocity vector} according to Rajaraman \cite{Rajaraman}. Use of
Lorentz invariance allows us to investigate the existence of some
kinds of solitary waves by reducing PDE (\ref{eq:nlkg}) to the
following ODE:
\begin{equation}
\frac{\partial^2 \phi_a}{\partial x^2}=\frac{\partial U}{\partial
\phi_a} \hspace{1cm} a=1,2, \dots , N \qquad . \label{eq:nlne}
\end{equation}
Therefore, the search for solitary waves or kinks, finite energy
solutions to the static field equations (\ref{eq:nlne}), is
tantamount to solving a analogous mechanical problem for a
unit-mass point particle moving in a plane with coordinates
$\phi=(\phi_1,\phi_2)$ under the influence of a potential
$-U(\phi)$, if the variable $x$ plays the r$\hat{\rm o}$le of \lq
\lq time" \cite{Rajaraman}.

The sine-Gordon and $\phi^4$ models, profusely dealt with in the
literature, are the basic examples of (\ref{eq:nlkg}),
respectively governed  by the PDE equations:
\[
\frac{\partial^2 \phi}{\partial t^2}-\frac{\partial^2
\phi}{\partial x^2}+\sin \phi=0 \hspace{2cm} \frac{\partial^2
\phi}{\partial t^2}-\frac{\partial^2 \phi}{\partial
x^2}+2\phi(1-\phi^2)=0 \qquad .
\]
The corresponding potential terms are $U(\phi)=1-\cos \phi$ and
$U(\phi)=\frac{1}{2}(\phi^2-1)^2$ and for both systems solitary or
traveling wave solutions exist: the well known sine-Gordon soliton
and $\lambda(\phi)^4_2$ kink. The peculiar non-dispersive
character of these non-linear waves is related to the structure of
the set ${\cal M}$ of zeroes of the potential $U(\phi)$. ${\cal
M}$ is a discrete set with more than one element. The field
profile of these solitary waves connects two elements of ${\cal
M}$ asymptotically. For instance, the kink $\phi_K(x)=\mp {\rm
tanh}x$ in the $\phi^4$ model interpolates between the two zeroes
$\phi_\pm=\pm 1$ of $U(\phi)$: $\phi_K(x=-\infty)=\pm 1$,
$\phi_K(x=\infty)=\mp 1$. Obviously, the greater the number of
elements in ${\cal M}$, the richer the solitary wave variety.
Thus, other models have been considered in the literature in the
search for these kinds of non-linear waves, such as the $\phi^6$
model with potential $U(\phi)=\frac{1}{2}\phi^2 (\phi^2-1)^2$, the
$\phi^8$ model with $U(\phi)=\frac{1}{2} (\phi^2-1)^2
(\phi^2-a^2)^2$ in the one-component scalar field theory framework
\cite{Lohe} and, more recently, polynomial interactions of any
order built from Chebyshev polynomials \cite{BGLM}.

Another way of obtaining a richer structure in the set ${\cal M}$,
and hence in the solitary wave manifold, is to increase the
dimension of the internal space, i.e, by considering theories with
more scalar field components. This is an important qualitative
step, as noted by Rajaraman \cite{Rajaraman}: \textit{This already
brings us to the stage where no general methods are available for
obtaining all localized static solutions, given the field
equations. However, some solutions, but by no means all, can be
obtained for a class of such Lagrangians using a little trial and
error}.

Straightforward generalization of the $\phi^4$ and $\phi^8$ models
to two-component scalar field theory leads to the potential energy
densities:
\begin{eqnarray}
v^{(A)}(\phi_1,\phi_2)&=&(\phi_1^2+\phi_2^2-1)^2 \label{eq:va}
\\
 v^{(B)}(\phi_1,\phi_2)&=&
(\phi_1^2+\phi_2^2-1)^2(\phi_1^2+\phi_2^2-a^2)^2 \label{eq:vb}
\qquad .
\end{eqnarray}
The action functional is invariant under $SO(2)$ rotations in the
$\phi_1-\phi_2$ internal plane. The zeroes of $v^{(A)}$ and
$v^{(B)}$, however, are not invariant under the $SO(2)$ action and
the orbits, ${\cal M}$, are continuous manifolds in these cases:
${\cal M}^{(A)}=S_1^{R=1}$, ${\cal M}^{(B)}=S_1^{R=1}\cup
S_1^{R=a}$, respectively. Goldstone bosons arise in the process of
quantification in this situation. Coleman proved in \cite{Co} that
there are no Goldstone bosons in a sensible scalar field theory on
the line. The infrared asymptotic behavior of quantum theory would
require the modification of the potentials $v^{(A)}$ and $v^{(B)}$
in such a way that their manifold of zeroes becomes a discrete
set. Thus, the effect of quantum fluctuations is to add
perturbations to the potentials (\ref{eq:va}) and (\ref{eq:vb})
\[
U(\phi_1,\phi_2)=v^{(A,B)}(\phi_1,\phi_2)+w(\phi_1,\phi_2)
\]
such that the $SO(2)$ symmetry is explicitly broken down into
discrete subgroups -acting on the new discrete sets of zeroes-
whereas the $U(\phi_1,\phi_2)$ potential energy density remains a
non-negative expression.

There are some models in the literature that match these features.
The best studied is the MSTB model, with potential energy density:
\[
U_{\rm  MSTB}(\phi_1,\phi_2)=v^{(A)}(\phi_1,\phi_2)+\sigma^2
\phi_2^2 \qquad ,
\]
where $\sigma$ is a non-dimensional parameter. It was first
proposed by Montonen \cite{Monto1} and Sarker, Trullinger and
Bishop \cite{Trulli1}. Rajaraman and Weinberg \cite{Raja1}
discovered two kinds of kinks by the trial orbit method; the TK1
(one-topological kinks) - tracing a straight line trajectory in
the analogous mechanical system- and the TK2 (two-component
topological kinks) - running through semi-elliptic orbits in the
mechanical system. These enquiries were followed by numerical
analysis \cite{Trulli2,Trulli3} and with this method Subbaswamy
and Trullinger found the existence of a whole family of
two-component non-topological kinks, which were named as NTK.
Moreover, they discovered an unexpected fact: the NTK energy is
equal to the addition of the TK1 and TK2 energies. This relation
is known as the kink mass \lq \lq {\it sum rule}". In 1985, Ito
explained all of these issues  analytically. The crux of the
matter is the separability of the Hamilton-Jacobi equation using
elliptic coordinates in the analogous mechanical system
\cite{Ito1,Ito2,J1,Aai2}. Unlike models with only one scalar
field, systems with two or more scalar fields have analogous
mechanical systems that  are generically non-integrable. The
analogous mechanical system of the MSTB model is a completely
integrable Type I Liouville model - separable in elliptic
coordinates- and all the solitary waves can be found analytically
. Other models exhibiting similar properties have been addressed
in References \cite{Mar1,Mar,Modeloa}. The generalization of this
kind of model to three-component scalar field theory is studied in
\cite{Aai3,Aai4,Aai5}.

The goal of this work is to identify the variety of solitary wave
or kink solutions in a broad family of models arising from
perturbations of the $SO(2)$ symmetric two-component scalar field
$\phi^4$ and $\phi^8$ models (\ref{eq:va}) and (\ref{eq:vb}). The
strategy will be to deal with analogous mechanical systems of
Liouville Type II, i. e., with Hamilton-Jacobi equations separable
in polar coordinates. These deformations necessarily involve a
singularity at the origin of the configuration space in the
mechanical system, equivalently, at the origin of the internal
plane in the field theoretical model. Strictly speaking, the
perturbation does not exist as a proper function when
$\phi_1=\phi_2=0$ because the limit of $w(\phi_1,\phi_2)$ when
both $\phi_1$ and $\phi_2$ are zero is either $0$ or $\infty$,
depending on the path followed to reach $w(0,0)$. This
mathematical pathology confers new and intriguing properties to
solitary wave solutions. For example, certain kink profiles
connecting identical vacuum points in internal space
asymptotically cannot be deformed into each other even though they
belong to the same topological sector in the configuration space.
Trajectories moving away from the origin involve infinite energy
and some kink solutions behave as strings pinned at one point in
internal space. Nevertheless, we shall show that in the four
models that we discussed here there is a plethora of kink or
solitary wave solutions, including topological and non-topological
kinks, one-parametric kink families, and singular solutions.

The organization of the paper is as follows: In Section \S 2 we
shall describe the models to be studied from a generic point of
view. We shall discuss the general properties of these models and
state three interesting results. In Sections \S 3, \S 4, \S 5 and
\S 6 we shall deal with the particular models that correspond to
two perturbations of (\ref{eq:va}) and (\ref{eq:vb}). Applying the
procedure explained in Section \S 2, we shall describe in detail
the solitary wave solutions arising in these systems and unveil
their hidden structure.

\section{Generalities}

In the models that we shall study, the (non-dimensional) scalar
fields,
\[
\vec{\phi}(x_0,x_1)=(\phi_1(x_0,x_1),\phi_2(x_0,x_1)):
\mathbb{R}^{1,1}\rightarrow \mathbb{R}^2
\]
are maps from (1+1)-dimensional Minkowskian space-time ${\mathbb
R}^{1,1}$ to the ${\mathbb R}^2$ internal space. The action
functional
\begin{equation}
S=\int d^2 x \left[\frac{1}{2} \sum_{a=1}^2\partial_\mu \phi^a
\partial^\nu \phi^a -U(\phi_1,\phi_2) \right] \label{eq:action}
\end{equation}
is invariant under the Poincar\'e transformations acting on
${\mathbb R}^{1,1}$, whereas the remaining symmetry
transformations of our models belong to the subgroup of $SO(2)$
rotations in the internal plane ${\mathbb R}^2$ that do not change
$U(\phi_1,\phi_2)$. Our convention for the metric tensor
components in Minkowski space ${\mathbb R}^{1,1}$ is
$g_{00}=-g_{11}=1$, $g_{12}=g_{21}=0$ and only non-dimensional
parameters will be considered throughout the paper.

A \lq\lq point" in the configuration space of the system is a
configuration of the field of finite energy; i.e., a picture of
the field at a fixed time such that the energy $E$, the integral
over the real line of the energy density,
\begin{equation}
 E[\vec{\phi}]=\int_{-\infty}^\infty dx {\cal E}[\vec\phi]\qquad ,
 \qquad {\cal
E}[\vec{\phi}]=\frac{1}{2} \left( \frac{d\phi_1}{d x} \right)^2
+\frac{1}{2} \left( \frac{d\phi_2}{d x} \right)^2
+U(\phi_1,\phi_2)  \label{eq:density}
\end{equation}
is finite. Thus, the configuration space is the set of continuous
maps from ${\mathbb R}$ to ${\mathbb R}^2$ of finite energy:
\[
{\cal C}=\left\{\vec{\phi}(x)\in {\rm Maps}({\mathbb R}, {\mathbb
R}^2)/E<\infty\right\} \qquad .
\]
In order to belong to ${\cal C}$, each configuration must comply
with the asymptotic conditions
\begin{equation}
\lim_{x\rightarrow \pm \infty} \vec{\phi} \in {\cal M}
\hspace{2cm} \lim_{x\rightarrow \pm \infty} \frac{d\vec{\phi}}{dx}
=0 \qquad , \label{eq:asymtotic}
\end{equation}
where ${\cal M}$ is the set of zeroes (minima) of the potential term
$U(\vec{\phi})$.

We shall analyze four different models, deforming the $\phi^4$ and
$\phi^8$ potential energy densities (\ref{eq:va}) and
(\ref{eq:vb}) by two classes of perturbations:
\begin{eqnarray}
 w^{(1)}(\phi_1,\phi_2)&=&\frac{\sigma^2 \phi_2^2}{(\phi_1^2+\phi_2^2)^2} \label{eq:w1} \\
 w^{(2)}(\phi_1,\phi_2)&=&
 \frac{\sigma^2}{2(\phi_1^2+\phi_2^2)}\left(1-
 \frac{\phi_1}{\sqrt{\phi_1^2+\phi_2^2}}
 \right) \label{eq:w2} \qquad ,
\end{eqnarray}
where $\sigma$ is a non-dimensional coupling constant of the
system. Thus, we consider four distinct potential energy densities
$U(\phi_1,\phi_2)$ labelled as follows:
\begin{equation}
U^{(IJ)}(\phi_1,\phi_2)=v^{(I)}(\phi_1,\phi_2)+w^{(J)}(\phi_1,\phi_2)
\label{eq:potii} \qquad ,
\end{equation}
where $I=A,B$ and $J=1,2$.

An important remark should be made here: both $w^{(1)}$ and
$w^{(2)}$ are singular at the origin. The limit of
$w^{(2)}(\phi_1,\phi_2)$ when $\phi_1\rightarrow 0$,
$\phi_2\rightarrow 0$ is always infinite, independently of the path
chosen in ${\mathbb R}^2$ to approach the origin.
$\lim_{(\phi_1,\phi_2)\rightarrow (0,0)} w^{(1)}(\phi_1,\phi_2)$ is,
however, finite, zero, if the origin is approached through the
abscissa axis $\phi_2=0$, but it is equal to infinity for any other
approaching path. Therefore, field profiles passing through the
origin have infinite energy and are excluded from the configuration
space in all of four cases, except field configurations such that
all their derivatives along the $\phi_2$ axis at the origin vanish
when the perturbation chosen is $w^{(1)}$.

The invariance of the system of ODE (\ref{eq:nlne}) under spatial
translations, $x\rightarrow x-x_0$ and reflections $x\rightarrow
-x$ makes it convenient to abbreviate the notation:
$\bar{x}=(-1)^\delta (x-x_0)$, $x_0\in\mathbb{R}$, $\delta=0,1$.
In this manner, we shall describe  a whole family of kinks and
anti-kinks with their centers located at arbitrary points on the
line, because these symmetry transformations bring solutions into
solutions.

The use of polar coordinates in the internal space ${\mathbb R}^2$
is suggested by the $SO(2)$ invariance of $v^{(A)}$ and $v^{(B)}$
as well as by the choice of perturbations $w^{(1)}$ and $w^{(2)}$.
This system of coordinates is defined by the diffeomorphism
\begin{eqnarray*}
&&\begin{array}{cccc} \zeta: & \mathbb{R}^+ \times S^1 &
\longrightarrow & \mathbb{R}^2
\\
& (R_0,\varphi_0)  & \longrightarrow & (\phi_1^0,\phi_2^0)
\end{array} \hspace{1cm} \left\{ \begin{array}{l} \zeta^*(\phi_1)=R
\cos \varphi
\\ \zeta^*(\phi_2)=R \sin \varphi \end{array} \right. \\
&& \quad 0\leq R=+\sqrt{\phi_1^2+\phi_2^2} \leq \infty \qquad
\qquad , \qquad \qquad 0\leq \varphi={\rm
arctan}{\phi_2\over\phi_1} < 2 \pi \qquad .
\end{eqnarray*}
By using polar coordinates, $v^{(A)}$, $v^{(B)}$, $w^{(1)}$ and
$v^{(2)}$ become:
\begin{eqnarray*}
&& f^{(A)}(R)=\zeta^* v^{(A)} = (R^2-1)^2
\hspace{1.cm};\hspace{1.2cm} f^{(B)}(R)=\zeta^* v^{(B)} =(R^2-1)^2
(R^2-a)^2 \hspace{0.2cm}; \\
&& {1\over R^2}g^{(1)}(\varphi)= \zeta^*
w^{(1)}=\frac{\sigma^2}{R^2}\sin^2 \varphi
\hspace{1.2cm};\hspace{1.2cm} {1\over R^2}g^{(2)}(\varphi)=
\zeta^* w^{(2)}=\frac{\sigma^2}{R^2}\sin^2 \frac{\varphi}{2}
\qquad .
\end{eqnarray*}
Thus, the first summand in (\ref{eq:potii}) is a function
$f^{(I)}(R)$ that depends only on the radial variable. The second
summand in (\ref{eq:potii}) is the product of a function
$g^{(J)}(\varphi)$ depending only on the angular variable times
the square of the inverse of $R$:
\begin{equation}
\zeta^* (U^{(IJ)})=f^{(I)}(R)+\frac{1}{R^2} g^{(J)}(\varphi)
\label{eq:potiirad} \qquad .
\end{equation}
We shall refer to this class of systems as Type II Liouville
models after the type of their analogous integrable mechanical
systems \cite{Pere}. We see the reason for the inevitability of
the singularity at $R=0$: the factor ${1\over R^2}$ in the second
summand of $U^{(IJ)}(R,\varphi)$. In $w^{(1)}$, however, the
singularity is not seen if the origin is reached through the paths
$\varphi=\pi$ {\underline {and}} $\varphi=0$. In the other case,
$w^{(2)}$, only passing through the origin via $\varphi=0$ kills
the singularity, but there is no continuous way out and the \lq\lq
particle" would become entrapped by the singularity.

The general structure of the set ${\cal M}$ of zeroes of the
potential energy density is easy to unveil: $\zeta^* (U)$ vanishes
if and only if the two summands on the right hand side of formula
(\ref{eq:potiirad}) are zero. Thus,
\begin{eqnarray*}
\zeta^* {\cal M}&=& \{ (R_i,\varphi_j)  \in \mathbb{R}^+ \times
S^1 \, / \, f^{(I)}(R_i)=0 \, \, , \, \, g^{(J)}(\varphi_j)=0 \,
\, , \, \, \, i=1,...,M \, \, , \, \, \, j=1,...,N\}
\end{eqnarray*}
and the $M\times N$ points $(R_i,\varphi_j)\in \zeta^* {\cal M}$
lie on the knots of a lattice where the laths are the two sets of
perpendicular straight lines:
\[
r_R=\{ R=R_i, i=1,...,M \} \hspace{1cm} \mbox{and} \hspace{1cm}
r_\varphi= \{ \varphi=\varphi_j , j=1,...,N \}\qquad .
\]
These lines are separatrix curves\footnote{Technically, these
curves are envelopes of separatrix trajectories of the analogous
dynamical system.}: there are no bounded trajectories crossing
these boundaries. In the field theoretical context, static
solutions are also enclosed in these domains.

Using polar coordinates, the energy functional reads:
\begin{equation}
\zeta^* E=\int \left[ \frac{1}{2} \left( \frac{dR}{dx} \right)^2+
\frac{1}{2} R^2 \left( \frac{d\varphi}{dx}
\right)^2+f(R)+\frac{1}{R^2} g(\varphi) \right]
\label{eq:acfu}\qquad .
\end{equation}
One can think of this functional in (\ref{eq:acfu}) either as the
static energy of the scalar field or, alternatively, as the action
functional for a particle of unit mass and $(R,\varphi)$ position
coordinates moving under the influence of a potential
$-\zeta^*{U}(R,\varphi)=-\left(f(R)+\frac{1}{R^2}
g(\varphi)\right)$ with evolution parameter $x$. Thus, the motion
equations of this analogous mechanical system are the ODE system:
\begin{equation}
\frac{d^2 R}{dx^2}=\frac{df(R)}{dR} \hspace{1cm};\hspace{1cm} R^2
\frac{d^2 \varphi}{dx^2}+\frac{dR}{dx}
\frac{d\varphi}{dx}=\frac{dg(\varphi)}{d\varphi} \label{eq:oderad}
\qquad .
\end{equation}

Lorentz invariance guarantees that finite action solutions of
(\ref{eq:oderad}) are traveling waves of the scalar field theory
with their center of mass located at the origin; Lorentz and
translation transformations provide all the solitary wave solutions
of the system obtained from the static ones by applying the
appropriate transformations.

To solve the ODE (\ref{eq:oderad}), we take advantage of the
existence of two independent first-integrals in the mechanical
system, namely:
\begin{equation}
I_1=\frac{1}{2} \left(\frac{dR}{dx} \right)^2+\frac{1}{2} R^2
\left(\frac{d\varphi}{dx} \right)^2-f(R)-\frac{1}{R^2} g(\varphi)
\hspace{0.8cm};\hspace{0.8cm} I_2=\frac{1}{2} R^4 \left(
\frac{d\varphi}{dx} \right)^2-g(\varphi) \label{eq:inte}\qquad .
\end{equation}
$I_1$ and $I_2$ are respectively the energy and the generalized
angular momentum in the analogous mechanical system. Written in
polar coordinates, the asymptotic conditions (\ref{eq:asymtotic})
guaranteeing finite energy in the field theory model read:
\begin{equation}
\lim_{x\rightarrow\pm \infty} (R(x),\varphi(x))
=(R_{i_\pm},\varphi_{j_\pm})\in \zeta^* {\cal M} \quad , \quad
\lim_{x\rightarrow\pm\infty}{dR\over dx}(x)=0 \quad , \quad
\lim_{x\rightarrow\pm\infty}{d\varphi\over dx}(x)=0
\label{eq:pasymtotic} \qquad .
\end{equation}
Thus, $I_1(x=\pm\infty)=0$ and $I_2(x=\pm\infty)=0$; because $I_1$
and $I_2$ are invariants of the mechanical system they vanish for
every point $x\in\mathbb R$. Therefore, solitary wave solutions of
scalar field theory are in one-to-one correspondence (modulo
Lorentz boosts) with the trajectories of the mechanical analogous
system such that $I_1=0$, $I_2=0$, i.e., trajectories solving the
following first-order ODE system:
\begin{equation}
\frac{dR}{dx}=(-1)^\alpha\sqrt{2f(R)} \qquad \qquad , \qquad
\qquad \frac{d\varphi}{dx}=(-1)^\beta {1\over
R^2}\sqrt{2g(\varphi)}  \quad , \quad \alpha,\beta=0,1 \quad .
\label{eq:fo2iz}
\end{equation}

We now state the first result concerning the structure of the
solitary wave variety in Type II Liouville models:

\vspace{0.1cm}

\textit{Proposition 1.-} There exist kink or solitary wave
solutions whose orbits lie on the separatrix straight lines. Kink
orbits are of two kinds: 1) those connecting two elements
$(R_i,\varphi_j)$ and $(R_{i+1},\varphi_j)$ of ${\cal M}$ through
a straight segment $(R(x),\varphi_j)$ (angular rays in the
Cartesian internal plane) and 2) those joining the vacuum points
$(R_i,\varphi_j)$ and $(R_i,\varphi_{j+1})$ through the straight
line $(R_i,\varphi(x))$ (circles on the Cartesian internal plane).

\vspace{0.1cm}

The proof is easy: note that the vanishing of the first integrals
$I_1=0, I_2=0$ along the kink orbits and the assumption of the
existence of solutions whose orbit is $R=R_i$ or
$\varphi=\varphi_j$ on the polar internal plane, $R_i$ and
$\varphi_j$ being respectively roots of the functions $f(R)$ and
$g(\varphi)$, are compatible conditions. We distinguish two
possible cases:
\begin{itemize}
\item If $R=R_i$,
\[
I_1=\frac{1}{2} R_i^2 \left( \frac{d\varphi}{dx}
\right)^2-\frac{1}{R_i^2} g(\varphi)=0=\frac{1}{R_i^2}I_2
\]
holds and the problem of finding the time-schedule or kink form
factor is solved by the quadrature:
\[
G[\varphi]=\int \, \frac{d\varphi}{\sqrt{2g(\varphi)}} \qquad ,
\qquad \varphi^K(x)=G^{-1}\left(\frac{\bar{x}}{R_i^2}\right) \qquad
.
\]

\item If $\varphi=\varphi_j$, the two invariants become
\[
I_1=\frac{1}{2} \left( \frac{dR}{dx} \right)^2-f(R)=0
\hspace{1cm}, \hspace{1cm} I_2=\frac{1}{2} g(\varphi_j)=0 \qquad .
\]
The vanishing of $I_2$ turns into a identity but the first-order
equation annihilating $I_1$ provides the kink form factor by
means of the quadrature:
\[
F[R]=\int \, \frac{dR}{\sqrt{2f(R)}} \qquad , \qquad
R^K(x)=F^{-1}(\bar{x}) \qquad .
\]
\end{itemize}

To obtain all the separatrix orbits -those for which $I_1=0$ and
$I_2=0$, in Type II Liouville models- it is convenient to apply
the Hamilton-Jacobi procedure because the HJ equation is separable
in polar coordinates and all the trajectories can be found by
quadratures. The generalized momenta are $p_R=\frac{dR}{dx}$ and
$p_\varphi=R^2 \frac{d\varphi}{dx}$, whereas the mechanical
Hamiltonian (or Hamiltonian energy density in the field theory)
is:
\[
{\cal H}=h_R+\frac{1}{R^2} h_\varphi \hspace{1cm} , \hspace{1cm}
h_R=\frac{1}{2} p_R^2-f(R) \hspace{1cm} , \hspace{1cm}
h_\varphi=\frac{1}{2} p_\varphi^2-g(\varphi) \qquad .
\]

The ansatz of separation of variables ${\cal J}={\cal
J}_R(R)+{\cal J}_\varphi(\varphi)-i_1 x$ for the Hamilton
principal function ${\cal J}$ converts the PDE Hamilton-Jacobi
equation
\[
\frac{\partial {\cal J}}{\partial x}+ {\cal H}\left(
\frac{\partial {\cal J}}{\partial R},\frac{\partial {\cal
J}}{\partial \varphi},R,\varphi\right)=0
\]
into the ODE system:
\begin{equation}
{1\over 2}\left(\frac{d{\cal J}_R}{dR}\right)^2-f(R)-{i_2\over
R^2}=i_1 \qquad , \qquad {1\over 2}\left(\frac{d{\cal
J}_\varphi}{d\varphi}\right)^2-g(\varphi)=-i_2 \quad ; \quad
i_1,i_2\in\mathbb R \label{eq:2inv}\qquad .
\end{equation}
Therefore, the Hamilton characteristic function is given by the
quadratures
\[
{\cal J}_R(R)= {\rm sign} (p_R) \, \sqrt{2} \int dR
\sqrt{f(R)+i_1+\frac{i_2}{R^2}} \qquad , \qquad {\cal
J}_\varphi(\varphi) ={\rm sign} (p_\varphi) \, \sqrt{2} \int
d\varphi \sqrt{g(\varphi)-i_2} \quad ,
\]
and the kink solutions comply with the equations:
\begin{itemize}
\item \lq\lq Orbit" equation: $ \left.\frac{\partial{\cal J}}{\partial
i_2}\right|_{i_1=0=i_2}=\gamma_1$,
\begin{equation}
{\rm sign} (p_R) \int \frac{dR}{R^2 \sqrt{2 f(R)}} - {\rm sign}
(p_\varphi) \int \frac{d\varphi}{\sqrt{2 g(\varphi)}} = \gamma_1
\label{eq:orbgen}
\end{equation}
\item \lq\lq Time schedule": $\left.\frac{\partial{\cal J}}{\partial i_1}
\right|_{i_1=0=i_2}=\gamma_2$,
\begin{equation}
{\rm sign} (p_R) \int \frac{dR}{\sqrt{2 f(R)}} = x+ \gamma_2
\qquad . \label{eq:espgen}
\end{equation}
\end{itemize}

To unveil how the special orbits $R=R_i$ and $\varphi=\varphi_j$
are hidden in the family of trajectories parametrized by
$\gamma_1$ and $\gamma_2$, it is useful to show explicitly the
connection between the systems of equations (\ref{eq:fo2iz}) and
(\ref{eq:orbgen})-(\ref{eq:espgen}). To fulfil this goal, we
derive from (\ref{eq:fo2iz}) three identities:
\begin{eqnarray}
\int \, \frac{dR}{\sqrt{2f(R)}}&=& \bar{x} \label{eq:fohj1}
\\ \int \, \frac{dR}{(-1)^\alpha R^2
\sqrt{2f(R)}}&=&\int \, \frac{dx}{R^2(x)}+c \label{eq:fohj2} \\
\int \, \frac{d\varphi}{(-1)^\beta\sqrt{2g(\varphi)}}&=&\int \,
\frac{dx}{R^2(x)}+d   \label{eq:fohj3}
\end{eqnarray}
It is clear that (\ref{eq:fohj1}) is equal to (\ref{eq:espgen}),
whereas subtracting (\ref{eq:fohj3}) from (\ref{eq:fohj2}) one
obtains (\ref{eq:orbgen}) if $\gamma_1=c-d$ and ${\rm
sign}(p_\varphi)=(-1)^\beta$. The straight line orbits arise when
$|\gamma_1|=\infty$. In this limit, the
(\ref{eq:orbgen})-(\ref{eq:espgen}) system makes sense only if (a)
$\varphi(x)=\varphi_j$ is the orbit, $g(\varphi_j)=0$, and the
time schedule (kink form factor or kink profile) is obtained by
integrating (\ref{eq:espgen}); (b) $R(x)=R_i$ is the orbit,
$f(R_i)=0$, and the time schedule is obtained by integrating
(\ref{eq:fohj3}). Thus, the special straight line orbits are at
the boundary of the family of kink trajectories.

Because the kink energy is the action of the associated separatrix
trajectory in the analogous mechanical system --$\left|{\cal
J}_R\right|+ \left|{\cal J}_\varphi\right|$ computed along the
kink path-- we state:

\vspace{0.1cm}

\textit{Proposition 2.-} The energy associated to a kink or solitary
wave solution in the Type II Liouville models is:
\[
{\cal E}[\phi(x)]= \left| \int_{\pi_r(\zeta^* \phi)} dR \sqrt{2
f(R)} \right|+  \left| \int_{\pi_\varphi(\zeta^* \phi)} d\varphi
\sqrt{2 g(\varphi)} \right| \qquad .
\]
Here, $\pi_R$ and $\pi_\varphi$ are the projectors onto the $R$-
and $\varphi$-axes in the polar cylinder $\mathbb R^+\times
\mathbb S^1$; application of these projectors to the
$x$-parametrized kink paths allows us to trade the path
integration along complicated curves by the sum of integrations
along straight $R(x)=R_i$ and $\varphi(x)=\varphi_j$ lines. This
is the rationale underlying the kink energy sum rules: all the
kinks or combinations of kinks having the same projections to the
polar axes carry the same energy.

\vspace{0.1cm}

To close this Section on generalities, we briefly describe how the
analogous mechanical system is related to supersymmetry. In
supersymmetric classical mechanics all the interactions are
derived from a superpotential $W(\phi_1,\phi_2)$ \cite{B3,B5},
related to the mechanical potential energy through the equation:
\begin{equation}
U(\phi_1,\phi_2)=\frac{1}{2} \left( \frac{\partial W}{\partial
\phi_1} \right)^2 + \frac{1}{2} \left( \frac{\partial W}{\partial
\phi_2} \right)^2 \label{eq:superpot} \qquad .
\end{equation}
This is no more than the Hamilton-Jacobi equation for $i_1=0$ of a
mechanical system with flipped potential energy
$V(\phi_1,\phi_2)=-U(\phi_1,\phi_2)$ (precisely as in the
analogous mechanical system), and the superpotential is tantamount
to the Hamilton characteristic function. The mechanical action
reads:
\[
E={1\over 2}\int \, dx \, \left(\frac{d\phi_1}{dx}
\frac{d\phi_1}{dx}+\frac{d\phi_2}{dx} \frac{d\phi_2}{dx}+
\frac{\partial W}{\partial\phi_1} \frac{\partial
W}{\partial\phi_1}+\frac{\partial W}{\partial\phi_2} \frac{\partial
W}{\partial\phi_2}\right)
\]
and this can be arranged ${\rm a}^\prime$ la Bogomolny
\cite{Bogo}:
\[
E={1\over 2}\int \, dx \,
\left[\left(\frac{d\phi_1}{dx}-\frac{\partial W}{\partial
\phi_1}\right)^2+\left(\frac{d\phi_2}{dx}-\frac{\partial W}{\partial
\phi_2}\right)^2\right]+ \int \, \left(d\phi_1\frac{\partial
W}{\partial \phi_1}+d\phi_2\frac{\partial W}{\partial \phi_1}\right)
\qquad .
\]
Thus, the Bogomolny bound $E_B=\left|\int \, dW \right|$ is
attained by solutions of the first-order differential equations:
\begin{equation}
\frac{d\phi_1}{dx}=\frac{\partial W}{\partial \phi_1}
\hspace{2cm}, \hspace{2cm} \frac{d\phi_2}{dx}=\frac{\partial
W}{\partial \phi_2} \label{eq:odesup} \qquad .
\end{equation}

Because a  first-order ODE system such as (\ref{eq:odesup}) is
easier to solve than second-order motion equations, it is
important to know when the PDE equation (\ref{eq:superpot}) is
solvable. Hamilton-Jacobi separable systems admit a complete
solution of such an equation and for Type II Liouville systems the
situation is:

\vspace{0.1cm}

\textit{Proposition 3}. Type II Liouville models admit four
superpotentials.

\vspace{0.1cm}

Proof. In polar coordinates (\ref{eq:superpot}) reads:
\begin{equation}
\frac{1}{2} \left[ \left( \frac{\partial W}{\partial R}
\right)^2+\frac{1}{R^2} \left( \frac{\partial W}{\partial \varphi}
\right)^2 \right]=f(R)+\frac{1}{R^2} g(\varphi) \label{eq:psuper}
\qquad .
\end{equation}
Searching for solutions of (\ref{eq:psuper}) such that
\begin{equation}
\frac{\partial^2 W}{\partial R \partial \varphi}=0 \qquad ,
\label{eq:tIIs}
\end{equation}
we plug the expression $W(R,\varphi)=F(R)+G(\varphi)$ into the
previous formula to find that $F$ and $G$ must satisfy the
differential equations:
\[
\frac{dF}{dR}=(-1)^\alpha \sqrt{2f(R)} \hspace{2cm} , \hspace{2cm}
\frac{dG}{d\varphi}=(-1)^\beta \sqrt{2g(\varphi)} \qquad ,
\]
with $\alpha,\beta=0,1$. Thus, the four superpotentials, the
complete solution of the HJ equation for mechanical energy equal to
zero, are the quadratures:
\[
W(R,\varphi)= (-1)^\alpha \int \! dR  \, \sqrt{2f(R)} +(-1)^\beta
\int \! d \varphi \, \sqrt{2g(\varphi)} \qquad .
\]
The associated first-order equations are our old friends
\begin{equation}
\frac{dR}{dx}=\frac{dF}{dR}=(-1)^\alpha \sqrt{2 f(R)} \hspace{2cm}
, \hspace{2cm}\frac{d\varphi}{dx}={1\over
R^2}\frac{dG}{d\varphi}=(-1)^\beta \frac{\sqrt{2g(\varphi)}}{R^2}
\label{eq:edobogo} \qquad .
\end{equation}
Note that there is the need of introducing a metric factor in
polar coordinates, which we have shown to be equivalent to
equations (\ref{eq:orbgen})-(\ref{eq:espgen}), obtained through
the Hamilton-Jacobi method. It is worthwhile mentioning that a
global change of sign in the superpotential trades the kink
$\phi_{\rm K}(x)$ for the antikink solutions $\phi_{\rm K}(-x)$.

A bonus of the use of the concept of superpotentials is that the
two invariants of Type II models can be written in Cartesian
coordinates in the unified way:
\begin{eqnarray*}
I_1&=& \frac{1}{2} \left( \frac{d\phi_1}{dx} \right)^2+\frac{1}{2}
\left( \frac{d\phi_2}{dx} \right)^2-\frac{1}{2} \left( \frac{d
W}{d\phi_1} \right)^2-\frac{1}{2} \left( \frac{d W}{d\phi_2}
\right)^2 \\
I_2&=& \frac{1}{2} \left( \phi_2 \frac{d\phi_1}{dx}- \phi_1
\frac{d\phi_2}{dx} \right)^2-\frac{1}{2} \left( \phi_2 \frac{d
W}{d\phi_1} -  \phi_1 \frac{d W}{d\phi_2} \right)^2 \qquad .
\end{eqnarray*}

The vanishing conditions $I_1=0$, $I_2=0$, required for the kink
or solitary wave trajectories, are guaranteed by two, rather than
one, systems of first-order ODE equations:
\begin{itemize}

\item One expected, equivalent to the ODE system (\ref{eq:odesup}):
\[
\frac{d\phi_1}{dx}=\frac{\partial W}{\partial \phi_1} \hspace{2cm}
, \hspace{2cm} \frac{d\phi_2}{dx}=\frac{\partial W}{\partial
\phi_2} \qquad .
\]

\item A new one, unexpected and awkward:
\begin{eqnarray}
\frac{d\phi_1}{dx}&=&\frac{\phi_1^2-\phi_2^2}{\phi_1^2+\phi_2^2}\frac{\partial
W}{\partial \phi_1} + \frac{2 \phi_1 \phi_2}{\phi_1^2+\phi_2^2}
\frac{\partial W}{\partial \phi_2}=F_1(\phi_1,\phi_2) \nonumber \\
\frac{d\phi_2}{dx}&=& \frac{2 \phi_1 \phi_2 }{\phi_1^2+\phi_2^2}
\frac{\partial W}{\partial
\phi_1}-\frac{\phi_1^2-\phi_2^2}{\phi_1^2+\phi_2^2}\frac{\partial
W}{\partial \phi_2}=F_2(\phi_1,\phi_2) \label{eq:2fo2iz} \qquad .
\end{eqnarray}
\end{itemize}
System (\ref{eq:2fo2iz}), however, can be written as the gradient
flow equations of a new superpotential $\tilde W (\phi)$:
\[
\frac{d\phi_1}{dx}=\frac{\partial{\tilde W}}{\partial \phi_1}
\hspace{2cm} , \hspace{2cm} \frac{d\phi_2}{dx}=\frac{\partial
{\tilde W}}{\partial \phi_2} \qquad .
\]
The reason is that $\frac{\partial
F_1}{\partial\phi_2}-\frac{\partial F_2}{\partial\phi_1}=0$, and
Green's theorem can be applied because the Type II separability
condition (\ref{eq:tIIs}) in Cartesian coordinates reads:
\[
-\phi_1\phi_2 \left( \frac{\partial^2 W}{\partial \phi_1 \partial
\phi_1}-\frac{\partial^2 W}{\partial \phi_2 \partial \phi_2}
\right)+(\phi_1\phi_1-\phi_2\phi_2) \frac{\partial^2 W}{\partial
\phi_1\partial \phi_2}+ \phi_1 \frac{\partial W}{\partial \phi_2}-
\phi_2 \frac{\partial W}{\partial \phi_1} =0 \qquad ,
\]
i.e., precisely the conditions necessary for the curl of the
vector field
$\vec{F}(\phi_1,\phi_2)=F_1(\phi_1,\phi_2)\vec{e}_1+F_2(\phi_1,\phi_2)\vec{e}_2$
being zero.

\section{The A1 Model}

We shall first choose $v^{(A)}$ and $w^{(1)}$ to build the
potential energy density:
\[
U^{(A1)}(\phi_1,\phi_2)=(\phi_1^2+\phi_2^2-1)^2+\frac{\sigma^2
\phi_2^2}{(\phi_1^2+\phi_2^2)^2} \qquad ,
\]
\begin{figure}[htb]
\centerline{\begin{tabular}{c}
\includegraphics[height=4.cm]{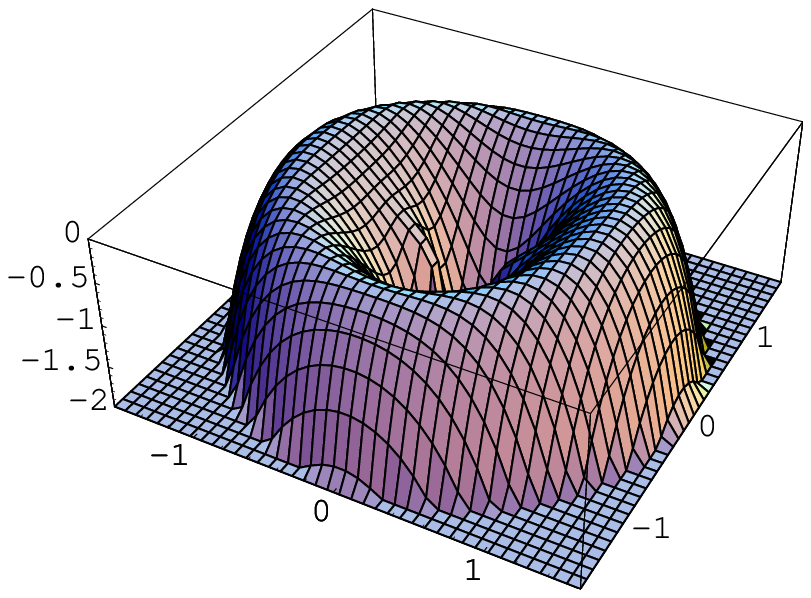}
\end{tabular}
\hspace{1cm}
\begin{tabular}{c}
\includegraphics[height=3.cm]{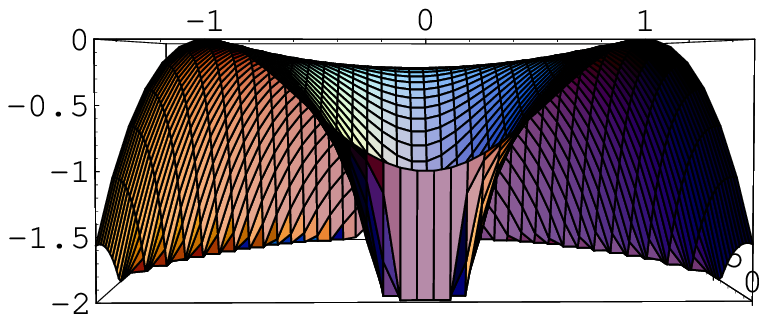}
\end{tabular}} \caption{\small \textit{Potential energy
$-U^{(A1)}(\phi_1,\phi_2)$ in the analogous mechanical system of
Model A1: a) Global perspective and b) Section showing the
singularity and the $\phi_2=0$ path.}}
\end{figure}
depicted in Figure 1. The second summand in the potential energy
density $U^{(A1)}(\phi_1,\phi_2)$ spoils the $SO(2)$ symmetry
preserved by the first summand. The manifold ${\cal M}$ of zeroes
of $U^{(A1)}(\phi_1,\phi_2)$ is a discrete set of two elements:
\[
{\cal M}=\{A_+=(1,0)\, ; \, A_-=(-1,0) \} \qquad .
\]
The symmetry group of this model is the vierergroup ${\mathbb
G}=\mathbb{Z}_2\times \mathbb{Z}_2$ generated by the reflections
$\phi_1 \rightarrow -\phi_1$ and $\phi_2\rightarrow -\phi_2$ in
the internal plane. This symmetry is spontaneously broken to the
${\mathbb Z}_2$ subgroup generated by $\phi_2\rightarrow -\phi_2$
through the choice of one of the two zeroes, because ${\cal M}$ is
the orbit of the other ${\mathbb Z}_2$ subgroup, in this case
generated by $\phi_1 \rightarrow -\phi_1$. The moduli space of
zeroes, the quotient space $\bar{\cal M}={\cal M}/{\mathbb G}=e$,
however, contains only one element $A=\{A_+,A_-\}$.

The partial differential field equations are:
\begin{eqnarray*}
\frac{\partial^2 \phi_1}{\partial t^2}-\frac{\partial^2
\phi_1}{\partial x^2}&=& 4 \phi_1 \left[1-\phi_1^2-\phi_2^2 +
\frac{\sigma^2 \phi_2^2}{(\phi_1^2+\phi_2^2)^3} \right] \\
\frac{\partial^2 \phi_1}{\partial t^2}-\frac{\partial^2
\phi_1}{\partial x^2}&=& \phi_2 \left[ 4-4 \phi_1^2-4 \phi_2^2
-\frac{2\sigma^2}{(\phi_1^2+\phi_2^2)^2}+\frac{4\sigma^2
\phi_2^2}{(\phi_1^2+ \phi_2^2)^3} \right] \qquad .
\end{eqnarray*}
In the search for solitary waves, however, we follow the general
procedure described in the previous Section because the analogous
mechanical system is a Type II Liouville model with:
\begin{equation}
f(R)=(R^2-1)^2  \hspace{2cm} , \hspace{2cm} g(\varphi)=\sigma^2
\sin^2 \varphi \label{eq:a1pot} \qquad \qquad .
\end{equation}
Back in Cartesian coordinates, the superpotential, obtained from
the solution of the Hamilton-Jacobi equations, reads:
\[
W[\phi_1,\phi_2]=\sqrt{2}\left\{(-1)^\alpha
\sqrt{\phi_1^2+\phi_2^2}\left[ \frac{1}{3} (\phi_1^2+\phi_2^2)-1
\right]-(-1)^\beta \frac{\sigma \phi_1}{\sqrt{\phi_1^2+\phi_2^2}}
\right\} \qquad .
\]

There are no finite action trajectories leaving the area bounded
by the circle $C_1 \equiv \phi_1^2+\phi_2^2=1$ of unit radius.
Also, finite action trajectories do not cross the radial ray
$r_1\equiv \phi_2=0$; for this reason we call these curves
separatrix curves{\footnote{Do not confuse with separatrix
trajectories, those arising for a particular choice of the
parameters of the Hamilton principal function lying at the
frontier between periodic and unbounded motion. Separatrix curves
are themselves separatrix trajectories in this sense, but very
special ones forming the envelop of the whole manifold of these
critical motions.}}.

\subsection{Solitary waves in the boundary of the kink moduli space}

From Proposition 1 we know that there exist singular solitary
waves whose orbits are $C_1$ or $R_1$. We use Rajaraman's trial
orbit method in order to identify the form factor of these
solitary wave solutions.

\vspace{0.1cm}

$\bullet$ ${\rm K1}_1^{AA^*}$: Plugging the orbit $C_1\equiv
\phi_1^2+\phi_2^2=1$ into the field equations the following solitary
wave solutions are found:
\begin{equation}
\vec{\phi}^{\,\,{\rm K1}_1^{AA^*}}(x)=  \tanh \sqrt{2} \sigma
\bar{x} \, \vec{e}_1\pm {\rm sech}\sqrt{2} \sigma \bar{x}\,
\vec{e}_2 \qquad . \label{eq:cirk}
\end{equation}
We refer to them as ${\rm K1}_1^{AA^*}$ using the notation
established in Reference \cite{Modeloa}; the subscript stands for
the number of lumps that the solitary wave is composed of - see
Figure 2- whereas the superscript specifies the elements of ${\cal
M}$ that are connected by the kink orbit in a generic way. $A$ can
be either $A_+$ or $A_-$ and by $A^*$ we mean the complementary
point in ${\cal M}$. In the case depicted in Figure 2, the spatial
dependence connects the points $A_+$ and $A_-$. The density energy
and the energy of these kink solutions are respectively
\[
{\cal E}^{{\rm K1}_1^{AA^*}}(x) = 2 \sigma^2 {\rm sech}^2
\sqrt{2}\sigma \bar{x}\hspace{1cm} , \hspace{1cm} E({\rm
K1}_1^{AA^*})=\left|W(A_-,0)-W(A_+,0)\right|=2 \sqrt{2} \sigma
\qquad .
\]
In Figure 2 the main features of the solution are depicted for the
choice of plus sign in (\ref{eq:cirk}) and $\bar x=x$: (a) the
kink form factor shows that both field components have non-zero
profiles. (b)  The kink orbits in the internal plane connect the
points in ${\cal M}$. (c) The energy density is localized around
one point and we interpret these solutions as basic solitary
waves. From a physical point of view they are seen as basic
traveling particles.

\begin{figure}[htb]
\centerline{\includegraphics[height=3.cm]{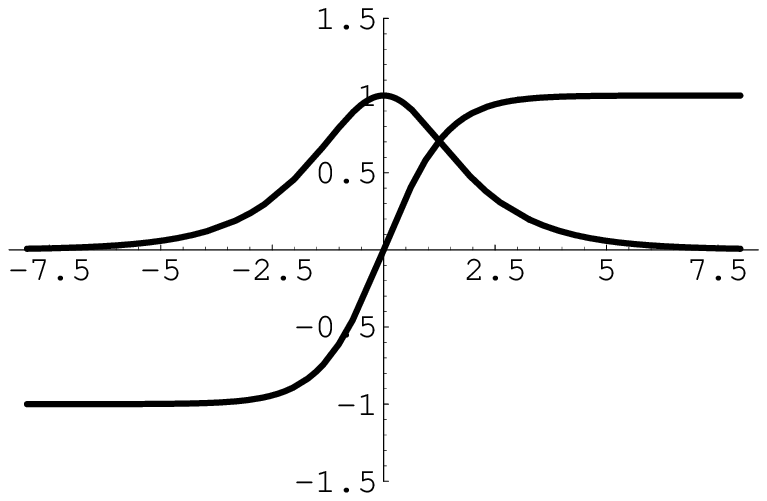}\hspace{1cm}
\includegraphics[height=3.cm]{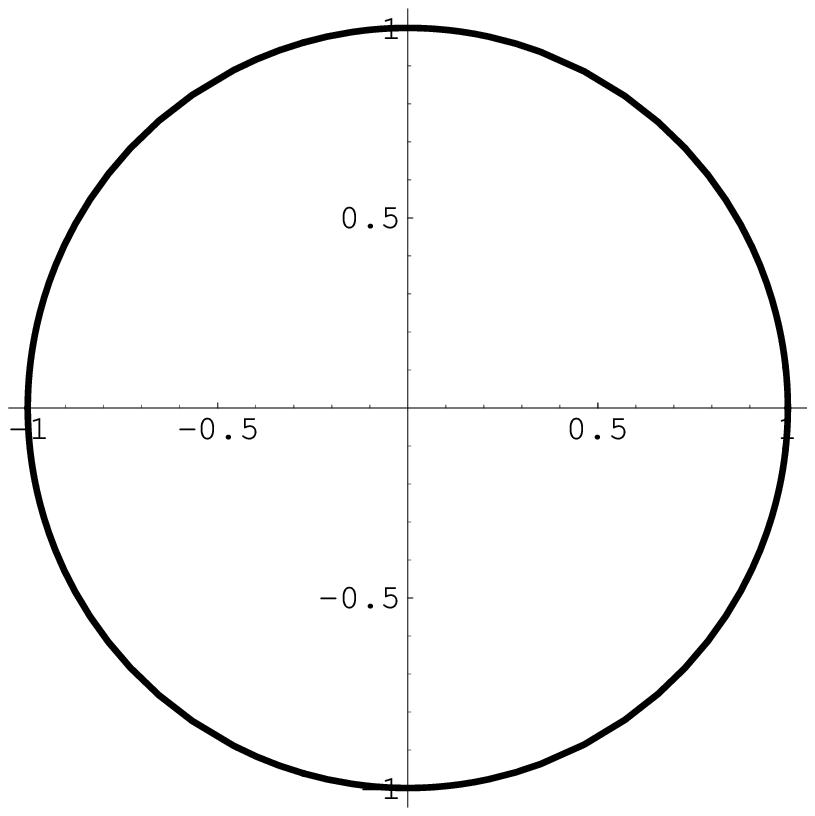}\hspace{1cm}
\includegraphics[height=3.cm]{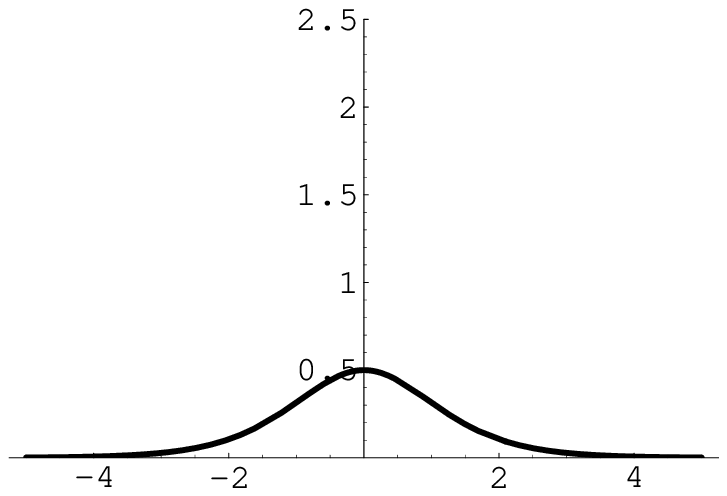}}
\caption{\small \textit{Solitary waves ${\rm K1}_1^{AA}$: a) Kink
form factor, b) Kink Orbits and c) Energy Density.}}
\end{figure}

\vspace{0.1cm}

$\bullet$ ${\rm K2}_1^{AA^*}$: Next, the $\phi_1$-axis $R_1\equiv
\phi_2=0$ is tried in the ODE system (\ref{eq:odesup}). The solution
\[
\vec{\phi}^{{\rm K2}_1^{AA^*}}(x)=\tanh \sqrt{2} \bar{x} \,
\vec{e}_1
\]
corresponds to solitary waves going from $A_-$ to $A_+$ as
$\bar{x}$ varies from $-\infty$ to $+\infty$. Note that these
kinks pass through the origin in the internal plane, a very
dangerous point. One could surmise that the singularity at this
point endows infinity energy to these solutions. We notice,
however, that these solutions follow the path $\phi_2=0$, where
the limit of $w^{(1)}(\phi_1,\phi_2)$ is zero, and these kinks do
not feel the singularity. In fact, the kink energy density
\[
{\cal E}^{{\rm K2}_1^{AA^*}}(x) = 2 \, {\rm sech}^4 \sqrt{2}\,
\bar{x}
\]
\[
E({\rm
K2}_1^{AA^*})=\left|W(A_-,\phi_2=0)-\lim_{\varepsilon\rightarrow
0^-}W(\varepsilon,\phi_2=0)\right|+\left|\lim_{\varepsilon\rightarrow
0^+}W(\varepsilon,\phi_2=0)-W(A_+,\phi_2=0)\right|=\frac{4\sqrt{2}}{3}
\]
is centered around one point in a more concentrated way that the
previous solutions - whereas the energy itself is different - and
${\rm K2}_1^{AA^*}$ kinks belong to another class of basic
particles in the model. We emphasize the following subtle point:
because $W(\phi_1,\phi_2)$ is a regular function all along the
$K1_1^{AA^*}$ kink orbit Stoke's theorem can be applied and
$E[K1_1^{AA^*}]$ depends only on the value of $W$ at the starting
and ending points $A_\pm$. The $K2_1^{AA^*}$ kink orbit, however,
hits the origin coming through the $\phi_2=0$ axis. The gradient
of $W$ is undefined in the neighborhood of the origin along this
path and careful application of Stoke's theorem gives the formula
for $E[K2_1^{AA^*}]$ written above. Note that $E[K1_1^{AA^*}]\geq
E[K2_1^{AA^*}]$ if $\sigma\geq {2\over 3}$ and $E[K1_1^{AA^*}]<
E[K2_1^{AA^*}]$ if $\sigma< {2\over 3}$. Figure 3 shows the main
features of one of these solitary waves, form factor, orbit and
energy density.

In order to gain a deeper understanding of the physical
consequences of passing the kink profile through the origin , we
shall analyze the stability of ${\rm K2}_1^{AA^*}$ kinks in some
detail. The second-order small fluctuation or Hessian operator
around these kinks is the diagonal matrix differential operator
\[
{\cal H}[{\rm K2}_1^{AA^*}]=\left( \begin{array}{cc} {\cal H}_{11}
& {\cal H}_{12}
\\ {\cal H}_{21} & {\cal H}_{22}
\end{array}\right)  =\left( \begin{array}{cc}
-\frac{d^2}{dx^2} -4 + 12 \frac{\sinh^2 \sqrt{2} x}{\cosh^2
\sqrt{2} x} & 0 \\ 0 & -\frac{d^2}{dx^2} -4 + 4 \frac{\sinh^2
\sqrt{2} x}{\cosh^2 \sqrt{2} x}  + 2 \sigma^2 \, \frac{{\rm
cosh}^4 \sqrt{2} x}{{\rm sinh}^4 \sqrt{2} x}
\end{array}\right) \,\, .
\]
The entry ${\cal H}_{11}$ rules the behavior of the tangent
perturbations to the kink orbit. ${\cal H}_{11}$ is an ordinary
Schr\"odinger operator of Posch-Teller type with $\lambda_0=0$ as
the lowest eigenvalue. The associated eigenfunction (or zero mode)
describes a perturbation that is a translation of the kink center,
see Figure 3(d). ${\cal H}_{22}$ regulates the orthogonal
fluctuations to the $\phi_2=0$ orbit. Here, the potential is a
positive definite infinite barrier with the singularity at
$\bar{x}=0\equiv\vec{\phi}^{{\rm K2}_1^{AA^*}}(0)=\vec{0}$. This
means that fluctuations pushing the kink profile away from the
origin cost infinite energy, see Figure 3(d). Therefore, the kink
profiles of this kind of solitary waves behave as strings pinned
at the origin of the internal plane. In sum, because the spectrum
of the Hessian operator is non-negative these solutions are
stable. Moreover, ${\rm K1}_1^{AA^*}$ and ${\rm K2}_1^{AA^*}$
kinks join the same vacuum points asymptotically and in models
without this kind of singularity they would live in the same
topological sector of the configuration space. In the present
system, however, they belong to different sectors of ${\cal C}$
because they cannot be homotopically deformed into each other
under the restriction of finite energy.

\begin{figure}[htb]
\centerline{\includegraphics[height=2.8cm]{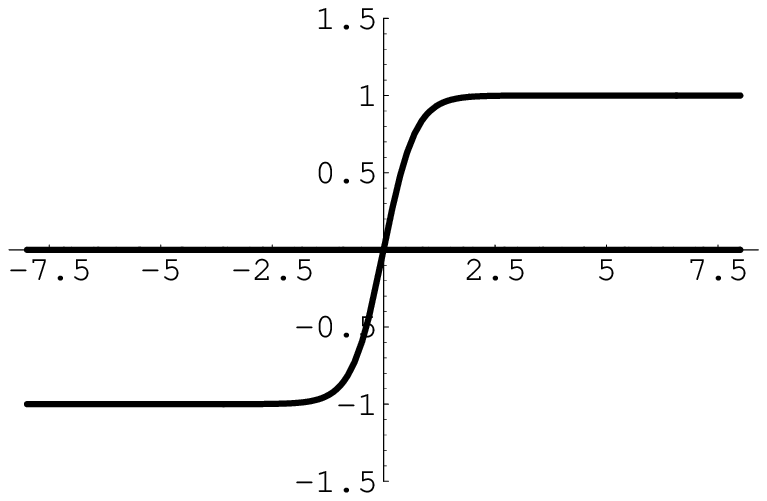}\hspace{0.1cm}
\includegraphics[height=2.8cm]{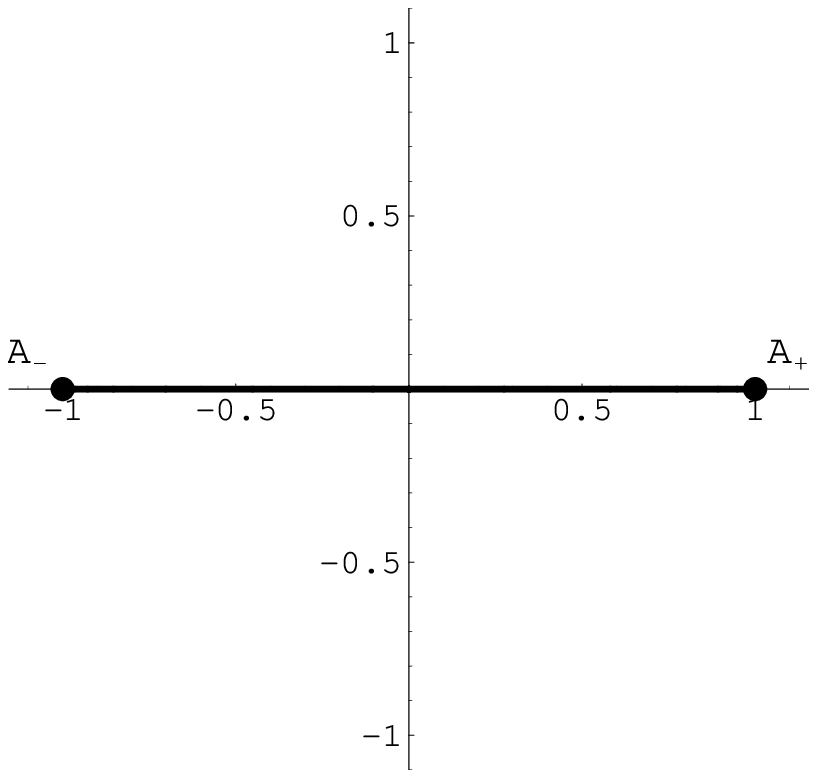}\hspace{0.1cm}
\includegraphics[height=2.8cm]{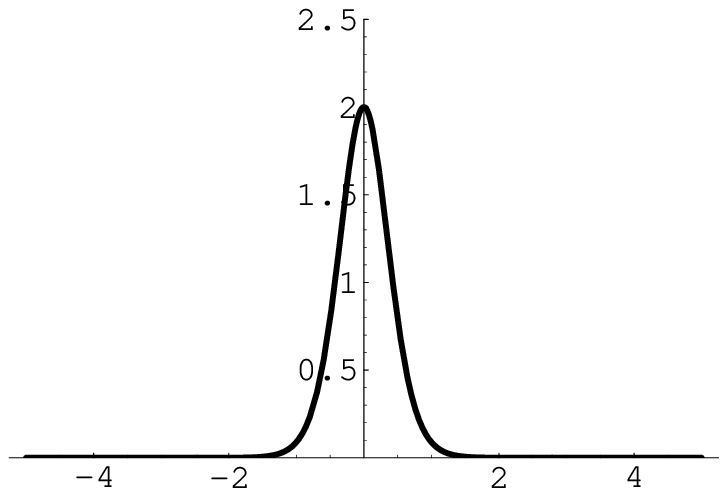}\hspace{0.1cm}
\includegraphics[height=2.5cm]{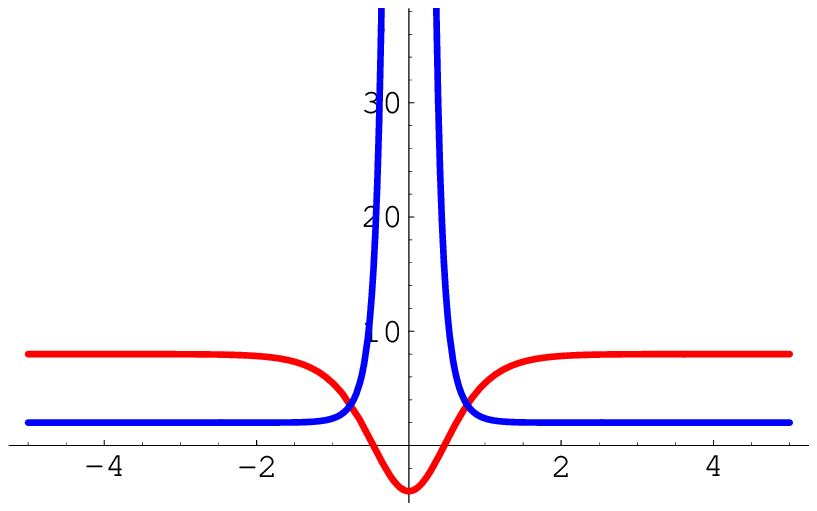}}
\caption{\small \textit{Solitary waves ${\rm K2}_1^{AA}$: a) Kink
form factor, b) Kink Orbits, c) Energy Density and d) Potential
Well and Barrier in the two diagonal matrix elements of the
Hessian operator.}}
\end{figure}

\subsection{Solitary waves in the bulk of the kink moduli space}

In this model, the Hamilton-Jacobi formulae (\ref{eq:orbgen}) and
(\ref{eq:espgen}), or equivalently, the Bogomolny first-order ODE
system (\ref{eq:edobogo}), become :
\begin{eqnarray}
(-1)^\alpha \int \, \frac{dR}{R^2\,(R^2-1)}-(-1)^\beta \int \,
\frac{d\varphi}{\sigma\,{\rm sin}\,\varphi}&=&\sqrt{2}
\label{eq:IIA1or1}\\\int \,
\frac{dR}{R^2-1}&=&\sqrt{2}\bar{x}\qquad . \label{eq:IIA1or2}
\end{eqnarray}
Integration of (\ref{eq:IIA1or1}) gives the kink orbits:
\begin{equation}
\log
\left(\left|\frac{R^K(x)+1}{R^K(x)-1}\right|^{\frac{(-1)^\alpha}{2}}\cdot
e^{-{1\over R^K(x)}}\right)-\log\left|{\rm
tan}\frac{\varphi^K(x)}{2}\right|^{\frac{(-1)^\beta}{\sigma}}=\sqrt{2}\gamma_1
\qquad .\label{eq:polo}
\end{equation}
The kink form factor (kink profile) in the $R$-variable is
obtained by integrating (\ref{eq:IIA1or2}):
\begin{equation}
R^K(x)={\rm tanh}\left(\sqrt{2}\left|\bar{x}
\right|\right)\label{eq:kppol}\qquad .
\end{equation}\vspace{0.1cm}
Plugging this solution into (\ref{eq:IIA1or1}), the kink profile
for the angular variable is found:
\begin{equation}
\varphi^K(x)=2{\rm arctan}\left[{\rm
exp}\left(\sqrt{2}\sigma(\bar{x}+\gamma_1)-\sigma{\rm
cotan}\bar{x}\right)\right] \qquad . \label{eq:kpol1}
\end{equation}

$\bullet$ ${\rm K}_3^{AA}(\gamma_1^-,\gamma_1^+)$:  In Cartesian
coordinates, these solutions -made out of two independent pieces-
read:
\begin{eqnarray*}
\vec{\phi}^{\,
K3^{AO}}(x)=\phi_1^{K3^{AO}}(x)\vec{e}_1+\phi_2^{K3^{AO}}(x)\vec{e}_2&=&
- \tanh \sqrt{2} |\bar{x}| \, \tanh \left[ \sigma (\sqrt{2}
(\bar{x}+\gamma_1)-{\rm cotanh}\, \sqrt{2} \bar x )\right]
\vec{e}_1
\\
&+& \,(-1)^\alpha \tanh \sqrt{2} |\bar{x}| \,\,{\rm sech} \left[
\sigma (\sqrt{2} (\bar{x}+\gamma_1)-{\rm cotanh}\, \sqrt{2} \bar x
)\right] \vec{e}_2 \qquad ,
\end{eqnarray*}
starting asymptotically from one of the minima denoted as $A$ and
and passing through the origin when $\bar{x}=0$. This point is
reached by each member of this family of solutions in such a way
that:
\[
\lim_{\bar{x}\rightarrow
0^-}\frac{d\phi_1}{d\bar{x}}=\sqrt{2}\hspace{1cm} , \hspace{1cm}
\lim_{\bar{x}\rightarrow 0^-} \frac{d^n \phi_2}{d\bar{x}^n}=0 \qquad
, \hspace{0.5cm} n=0,1,2,\dots \qquad .
\]
This point is crucial, because every solution of this type can be
continuously glued with any solution leaving the origin with the
same tangency properties and ending at the other point of the
vacuum orbit. The whole solution on the real line is thus a kink
avoiding the singularity, in the same way as the $K2_1^{AA}$ kink
does. In sum, there is a two-parametric family of kinks because we
can freely choose HJ trajectories with different integration
constants $\gamma_1^-$ and $\gamma_1^+$ in different regions:
$\bar{x}<0$ or $\bar{x}>0$.

A special member of this family is the case when
$\bar{\gamma}_1=-\infty$:
\begin{equation}
\vec{\phi}^{{\rm K}_2^{AA}(\gamma_1,-\infty)}(x)= \left\{
\begin{array}{cll}
\begin{array}{l}  - \tanh \sqrt{2}\, |\bar{x}|
\, \tanh [ \sigma (\sqrt{2} (\bar{x}+\gamma_1)-{\rm cotanh}\,
\sqrt{2}\, \bar x )]\, \vec{e}_1 \,+ \\ +\, (-1)^\alpha \tanh
\sqrt{2}\,  |\bar{x}| \,\, {\rm sech} [ \sigma (\sqrt{2}
(\bar{x}+\gamma_1)-{\rm cotanh}\, \sqrt{2} \,\bar x
)] \, \vec{e}_2 \end{array} & \mbox{ if } & \bar{x} \leq 0 \\[0.5cm] \tanh \sqrt{2}\,
\bar{x} & \mbox{ if } & \bar{x}>0
\end{array} \right. \label{eq:a1k2}\qquad ,
\end{equation}
which is formed by gluing a kink in the bulk determined by a
finite value of $\gamma_1^-$ at the left of the origin with a kink
in the boundary at the right of the origin. Note that for this
kind of solution the Hamilton-Jacobi procedure works independently
on the left and right half-lines and this is the reason for the
dependence on $\gamma_1^-$ and $\gamma_1^+$. In Figure 4 we depict
the kink field profiles for an $(\gamma_1^-,-\infty)$ orbit with
$\bar{x}=0$ and $\alpha=0$. Several kink orbits are also drawn to
show that the peculiarity of choosing $\gamma_1^+=-\infty$ changes
the ending point: these kink orbits link the point $A$ of the
vacuum orbit with itself by means of a trajectory crossing the
origin.
\begin{figure}[htb]
\centerline{\includegraphics[height=3.cm]{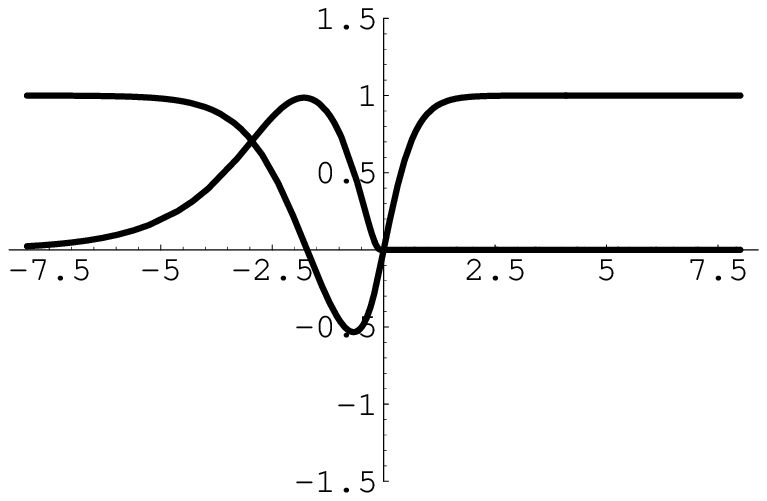}\hspace{1cm}
\includegraphics[height=3.cm]{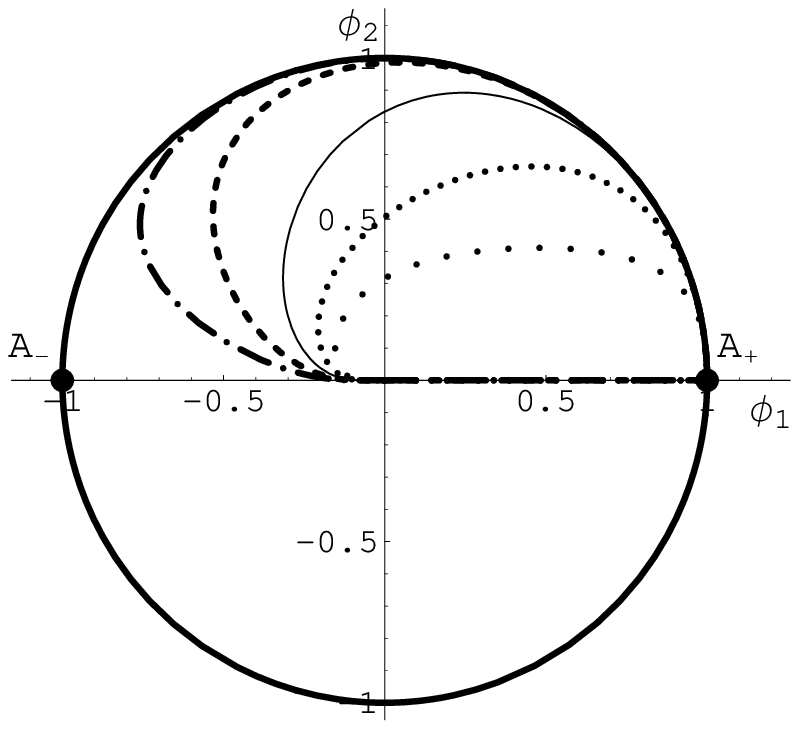}}
\caption{\small \textit{Solitary wave family ${\rm
K}_2^{AA}(\gamma_1^-,-\infty)$: a) Kink form factor and b) Orbits
for the values $\gamma_1^-=0,\pm 1,\pm 2$.}}
\end{figure}

Moreover, a Mathematica plot of the energy density for several
values of $\gamma_1^-$ (see Figure 5) shows that these solutions
are indeed a non-linear superposition of the basic solitary waves
discussed above: one ${\rm K1}_1^{AA}$ kink plus one ${\rm
K2}_1^{AA}$ kink. Observe, however, that the solutions
parametrized by $\gamma_1^-$ are composed of one $K1_1^{AA}$ and
one-half $K2_1^{AA}$ kinks whereas the choice of
$\gamma_1^+=-\infty$ continuously glues the remaining one-half
$K2_1^{AA}$ kink. This justifies the nomenclature ${\rm
K}_2^{AA}(\gamma_1^-,-\infty)$ for this kink family. If
$\gamma_1^-$ is positive, these solutions behave as two separate
lumps whereas if $\gamma_1^-$ is negative the two lumps sit on top
of each other.
\begin{figure}[htb]
\centerline{\includegraphics[height=2.3cm]{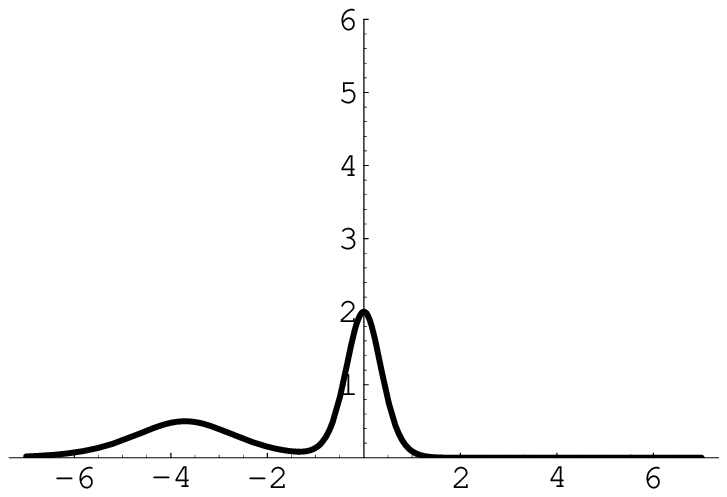}
\includegraphics[height=2.3cm]{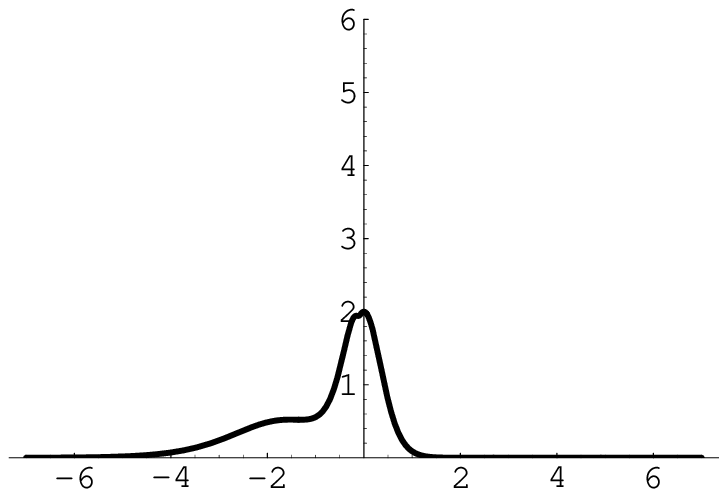}
\includegraphics[height=2.3cm]{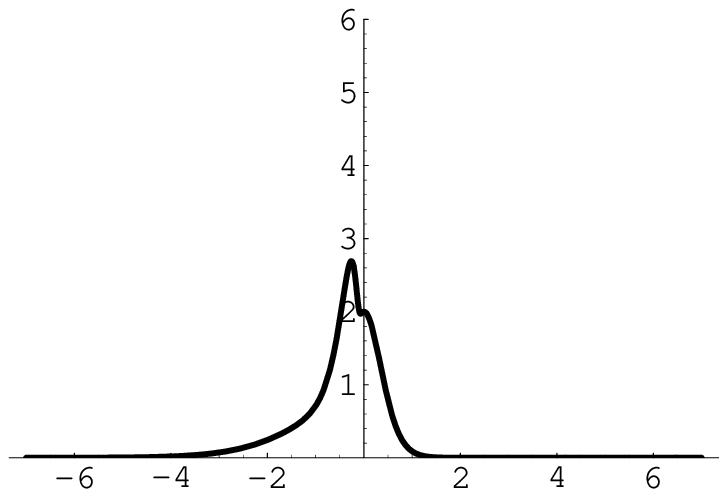}
\includegraphics[height=2.3cm]{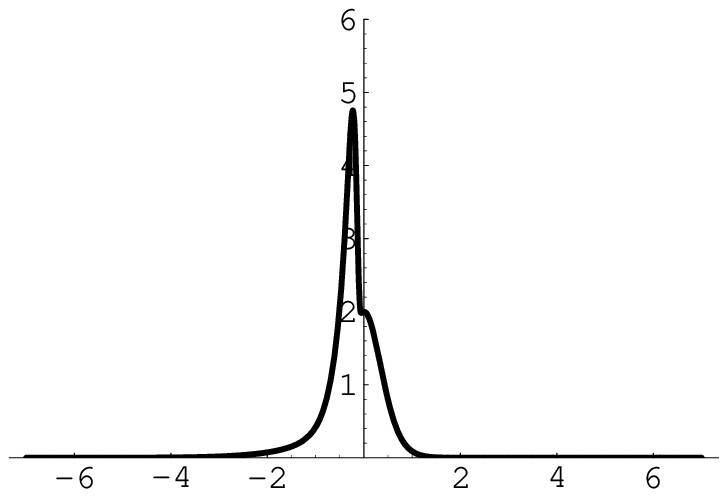}}
\caption{\small \textit{Energy density of the ${\rm
K}_2^{AA}(\gamma_1,-\infty)$ solitary waves  for decreasing values
of $\gamma_1^-$.}}
\end{figure}

The energy of these solutions is evaluated using expression
(\ref{eq:density}) to find: $E[{\rm
K}_2^{AA}(\gamma_1^-,-\infty)]=\frac{4\sqrt{2}}{3}+2\sqrt{2}
\sigma$. The following kink energy sum rule
\[
E[{\rm K}_2^{AA}(\gamma_1^-,-\infty)] = E[{\rm K1}_1^{AA}] +
E[{\rm K2}_1^{AA}]
\]
holds, showing the composite character of these solitary waves
built from two basic kinks.

Fully general solutions have the form:
\[
\vec{\phi}^{{\rm K}_3^{AA^*}(\gamma_1^+,\gamma_1^-)}(x)=\left\{
\begin{array}{l} \phi_1^{K_3^{AO}}(\bar{x};\gamma_1^-)\,
\vec{e}_1+\phi_2^{K_3^{AO}}(\bar{x};\gamma_1^-) \, \vec{e}_2 \mbox{ if } \bar{x} \leq 0\\
\phi_1^{K_3^{AO}}(\bar{x};\gamma_1^+)\,
\vec{e}_1-\phi_2^{K_3^{AO}}(\bar{x};\gamma_1^+) \, \vec{e}_2
\mbox{ if } \bar{x}
> 0 \end{array} \right. \qquad .
\]
Their energy density is the sum of two ${\rm K1}_1^{AA}$ and one
${\rm K2}_1^{AA}$ lumps, and is therefore a composite of three
basic kinks:
\[
E[{\rm K}_3^{AA^*}]=2 E[{\rm K1}_1^{AA^*}]+E[{\rm
K2}_1^{AA^*}]=4\sqrt{2} \sigma + \frac{4\sqrt{2}}{3}
\]

\begin{figure}[htb]
\centerline{\includegraphics[height=2.3cm]{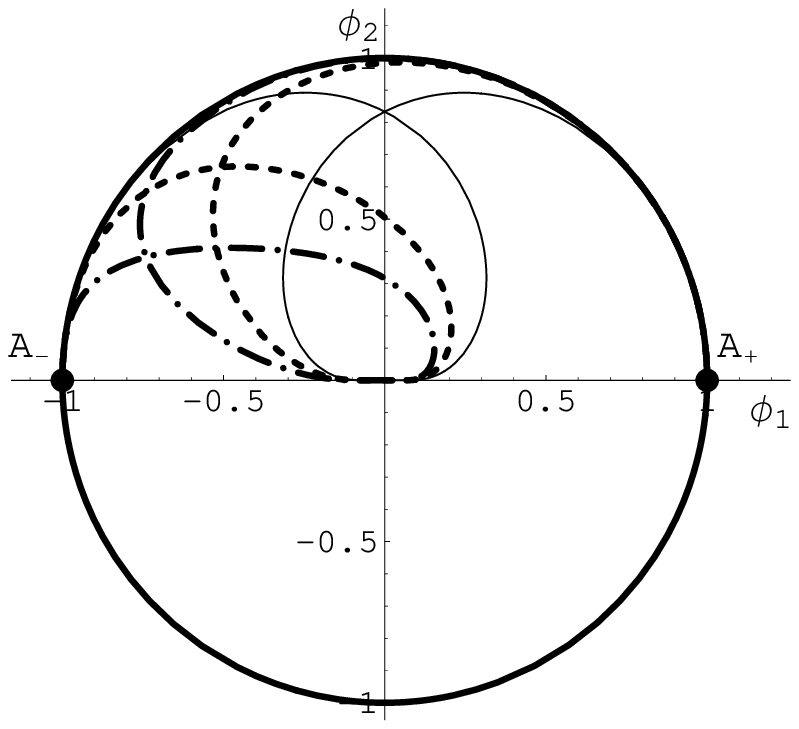}\hspace{1cm}
\includegraphics[height=2.3cm]{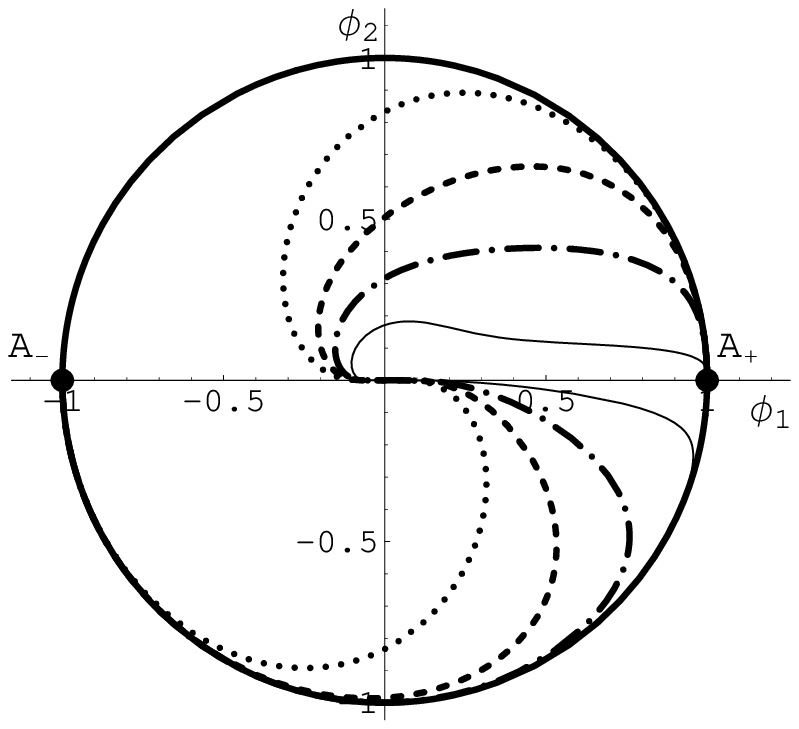}} \caption{\small \textit{Several $K_3^{AA}(\gamma_1^+,\gamma_1^-)$ kink
orbits}}
\end{figure}

A short remark on stability: $K_2^{AA}(\gamma_1^-,-\infty)$
solitary waves are non-topological whereas the
$K_3^{AA}(\gamma_1^-,\gamma_1^+))$ family is formed by topological
kinks. Both types, however, are stable because, passing through
the origin, they cannot decay to lighter kinks with the same
asymptotic behavior for identical reasons that forbid the decay of
the $K2_1^{AA}$ to the $K1_1^{AA}$ kink.

\section{The B1 Model }

In this section we study the deformation induced on the potential
$v^{(B)}(\phi_1,\phi_2)$ (\ref{eq:vb}) by $w^{(1)}(\phi_1,\phi_2)$
(\ref{eq:w1}). The dynamics is governed by the action functional
(\ref{eq:action}), with potential energy density:
\[
U^{(B1)}(\phi_1,\phi_2)=(\phi_1^2+\phi_2^2-1)^2(\phi_1^2+\phi_2^2-a^2)^2+\frac{\sigma^2
\phi_2^2}{(\phi_1^2+\phi_2^2)^2} \qquad .
\]
Thus, as in the $A1$ model, by adding the $w^{(1)}$ perturbation
the symmetry under the $SO(2)$ group is explicitly broken to the
discrete subgroup ${\mathbb G}=\mathbb{Z}_2\times \mathbb{Z}_2$
generated by reflections of the fields:  $\phi_1 \rightarrow
-\phi_1$, $\phi_2 \rightarrow -\phi_2$.

For the sake of simplicity we set the value of $a$ to be:  $a=2$.
Other non-null values of this parameter yield the same qualitative
behavior of the system with only analytic differences in the
specific expressions. The set of zeroes of
$v^{(B)}(\phi_1,\phi_2)$ is a continuous manifold with two
connected components: the disjoint union of two circles of radius
$1$ and $2$, $S^1_{r=1}\cup S^1_{r=2}$. The $w^1$ perturbation,
however, forces the set of zeroes of $U^{(B1)}$ to be the discrete
set of four points:
\[
{\cal M}=\{ B^-=(-2,0) \, \, ; \, \,  A^-=(-1,0) \, \, ; \, \, A^+=
(1,0) \, \, ; \, \, B^+=(2,0) \} \qquad .
\]
The vacuum orbit is the union of two ${\mathbb G}$-orbits: ${\cal
M}=\{A^-,A^+\}\cup \{B^-,B^+\}$. The moduli space of vacua
$\bar{\cal M}={\cal M}/{\mathbb G}=A \cup B$, however, is the union
of only two elements: $A=\{A^-,A^+\}$, $B=\{B^-,B^+\}$.

The field equations are:
\begin{eqnarray*}
\frac{\partial^2 \phi_1}{\partial t^2}-\frac{\partial^2
\phi_1}{\partial x^2}&=& 4 \phi_1\left[ (1-\phi_1^2-\phi_2^2)(4-
\phi_1^2-\phi_2^2)(5-2\phi_1^2-2\phi_2^2)+
\frac{\sigma^2 \phi_2^2}{(\phi_1^2+\phi_2^2)^3} \right] \\
\frac{\partial^2 \phi_2}{\partial t^2}-\frac{\partial^2
\phi_2}{\partial x^2}&=& 4 \phi_2 \left[ (1-\phi_1^2-\phi_2^2)(4-
\phi_1^2-\phi_2^2)(5-2\phi_1^2-2\phi_2^2)-\frac{\sigma^2}{2(\phi_1^2+\phi_2^2)^2}+
\frac{\sigma^2 \phi_2^2}{(\phi_1^2+\phi_2^2)^3} \right] \qquad .
\end{eqnarray*}
 The analogous mechanical system is the Type II Liouville system
 for which:
\[
f(R)=(1-R^2)^2 (4-R^2)^2 \hspace{2cm} , \hspace{2cm}
g(\varphi)=\sigma^2 \sin^2 \varphi \qquad .
\]
Back in Cartesian coordinates, the superpotentials, obtained from
the solution of the Hamilton-Jacobi equation, read:
\[
W[\phi_1,\phi_2]= \sqrt{2}\left\{
(-1)^\alpha\sqrt{\phi_1^2+\phi_2^2}\left[ \frac{1}{5}
(\phi_1^2+\phi_2^2)^2-\frac{1+a^2}{3}(\phi_1^2+\phi_2^2)+a^2
\right]-(-1)^\beta \frac{\sigma \phi_1}{\sqrt{\phi_1^2+\phi_2^2}}
\right\} \qquad .
\]
There are no finite action trajectories trespassing the circle:
$C_2\equiv \phi_1^2+\phi_2^2=4$. Also, $C_1 \equiv
\phi_1^2+\phi_2^2=1$, and $R_1\equiv \phi_2=0$ are not crossed by
any finite-action trajectories. Thus, $C_1$, $C_2$, and $R_1$ form
the set of separatrix curves of the $B1$ model.

\subsection{Solitary waves at the boundary of the moduli space}

From Proposition 1 in section \S 2, the existence of singular
solitary waves with  $C_2$, $C_1$, and $R_1$ orbits follows. Use
of the Rajaraman´s trial orbit method is effective.

$\bullet$ $K1_1^{AA^*}$. Substituting $\phi_1^2+\phi_2^2=1$ into
(\ref{eq:edobogo}) we obtain the following solitary wave solutions:
\[
\vec{\phi}^{\,{\rm K1}_1^{AA^*}}(x)=  \tanh \sqrt{2} \sigma \bar{x}
\, \vec{e}_1\pm  \,{\rm sech} \sqrt{2} \sigma \bar{x} \, \vec{e}_2
\qquad ,
\]
joining the points $A_+$ and $A_-$, see Figure 7. They are basic
lumps or particles of the $B1$ model - recall that exactly the
same solutions are also solutions of model $A1$- and their energy
density is concentrated around a point in the real line, see
Figure 7c. Specifically, we find that the energy density and the
energy of these kinks is exactly the same as in model $A1$:
\[
{\cal E}^{{\rm K1}_1^{AA^*}}(x) = 2 \sigma^2 {\rm sech}^2 \sqrt{2}
\sigma \bar{x} \hspace{1cm} , \hspace{1cm} E[{\rm
K1}_1^{AA^*}]=\left|W(A_-,0)-W(A_+,0)\right|=2\sqrt{2} \sigma \qquad
.
\]
\begin{figure}[htb]
\centerline{\includegraphics[height=3.cm]{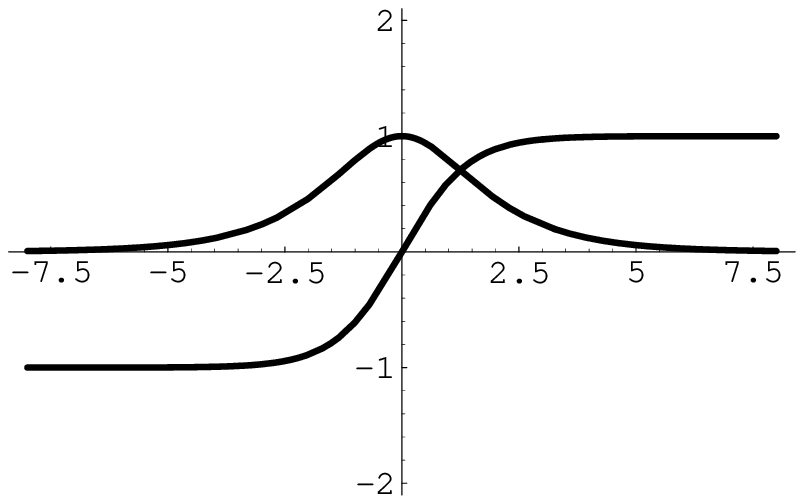}\hspace{1cm}
\includegraphics[height=3.cm]{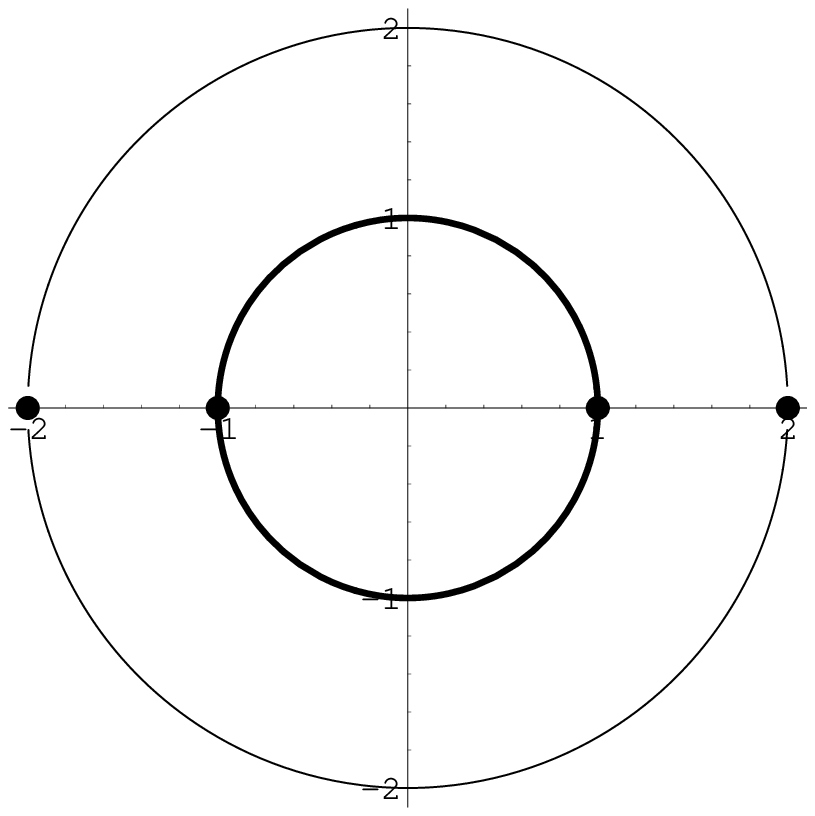}\hspace{1cm}
\includegraphics[height=3.cm]{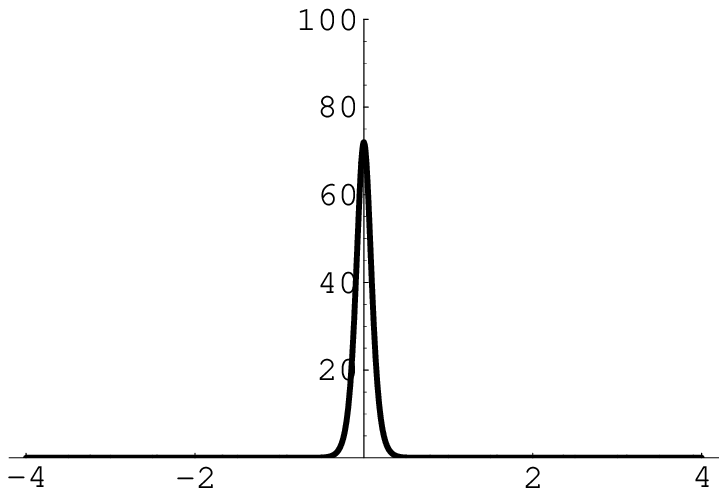}}
\caption{\small \textit{Solitary waves ${\rm K1}_1^{AA^*}$: a)
Factor Form, b) Orbit and c) Energy Density.}}
\end{figure}

\vspace{0.1cm}

$\bullet$ $K_1^{BB^*}$. Analogously, we choose
$\phi_1^2+\phi_2^2=4$ as the trial orbit and plug this expression
into (\ref{eq:edobogo}), to obtain:
\[
\vec{\phi}^{\,{\rm K}_1^{BB^*}}(x)=  2 \tanh \frac{\sigma
\bar{x}}{2\sqrt{2}} \, \vec{e}_1+2 \,{\rm sech} \frac{\sigma
\bar{x}}{2\sqrt{2}}\, \vec{e}_2 \qquad .
\]
These solitary waves share similar features with the $K_1^{AA^*}$
kinks, although in this case the asymptotically linked points are
$B_+$ and $B_-$. Even though the total amount of energy carried
out by the ${\rm K}_1^{BB^*}$ and ${\rm K1}_1^{AA^*}$ kinks is the
same, the energy density is less concentrated in the ${\rm
K}_1^{BB^*}$ kinks:
\[
{\cal E}^{{\rm K1}_1^{BB^*}}(x) = {\sigma^2\over 2} {\rm sech}^2
{\sigma \bar{x}\over 2\sqrt{2}} \hspace{1cm} , \hspace{1cm} E[{\rm
K1}_1^{BB^*}]=\left|W(B_-,0)-W(B_+,0)\right|=2\sqrt{2} \sigma \qquad
,
\]
see Figure 8. They describe different basic particles, living in
different topological sectors.
\begin{figure}[htb]
\centerline{\includegraphics[height=3.cm]{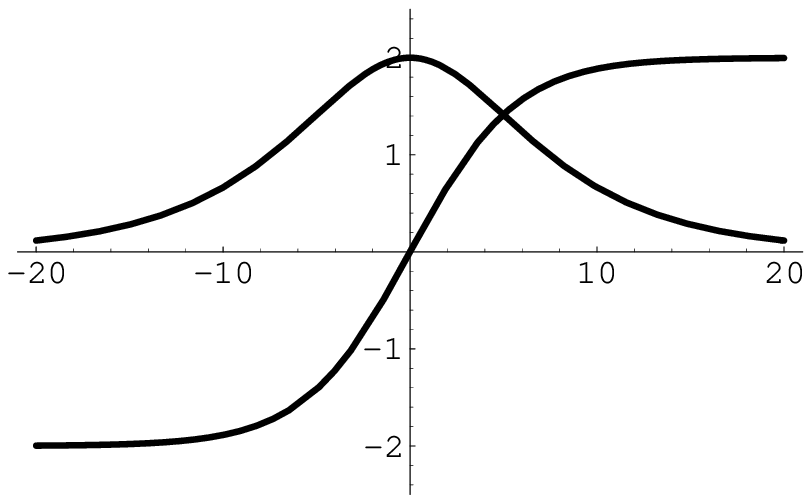}\hspace{1cm}
\includegraphics[height=3.cm]{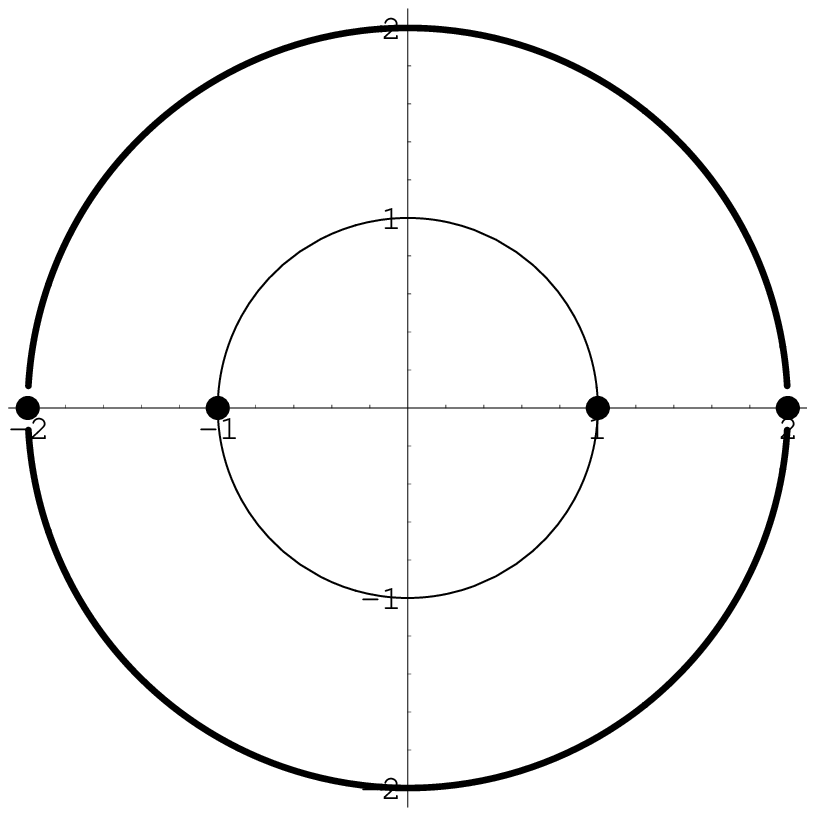}\hspace{1cm}
\includegraphics[height=3.cm]{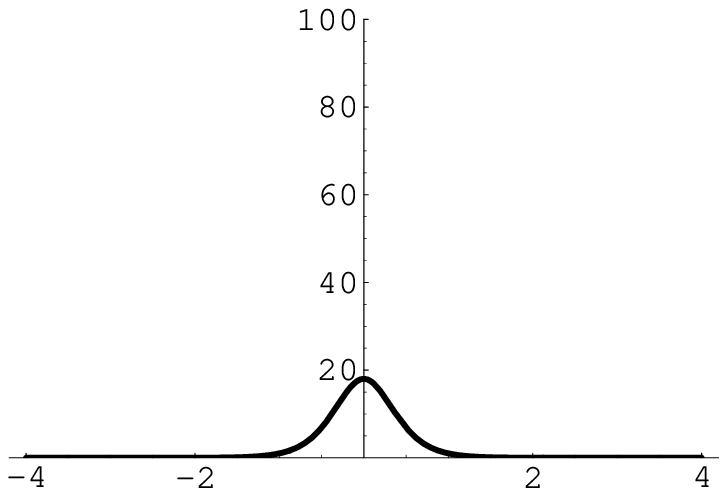}}
\caption{\small \textit{Solitary waves ${\rm K}_1^{BB^*}$: a)
Factor Form, b) Orbit and c) Energy Density.}}
\end{figure}

\vspace{0.1cm}

On the $\phi_2=0$  axis there are two different kinds of singular
solitary waves.

$\bullet$ $K_1^{AB}$:  The first kind includes two kink orbits,
the closed intervals $[B_-,A_-]$ -$\aleph=1$- and $[A_+,B_+]$
-$\aleph=0$-. The corresponding kinks connect points in different
elements $A$ and $B$ of the vacuum moduli space. We denote this
kind of solitary wave solutions as $K_1^{AB}$ kinks and their
profiles or form factors are easily found to be:
\[
\vec{\phi}^{\,{\rm K}_1^{AB}}(x)= 2 \cos \left[ \frac{2}{3} \arctan
e^{6\sqrt{2} \bar{x}}+\frac{2\pi}{3} \aleph\right] \vec{e}_1 \qquad
, \qquad \aleph=0,1
\]
The energy density and total energy of these solutions is:
\[
{\cal E}^{{\rm K}_1^{AB}}(x) = 32 \, {\rm sech}^2 6 \sqrt{2} \bar{x}
\, \sin^2 \left( \frac{2}{3} \arctan e^{6 \sqrt{2}
\bar{x}}+{2\pi\over 3}\aleph \right)
\]
\[
E[{\rm
K}_1^{AB}]=\left|W(A_\mp,0)-W(B_\mp,0)\right|=\frac{22\sqrt{2}}{15}
\qquad .
\]
The energy density is again localized around a point and these
solutions are also basic solitary waves or lumps of the $B1$
model, see Figure 9.
\begin{figure}[htb]
\centerline{\includegraphics[height=3.cm]{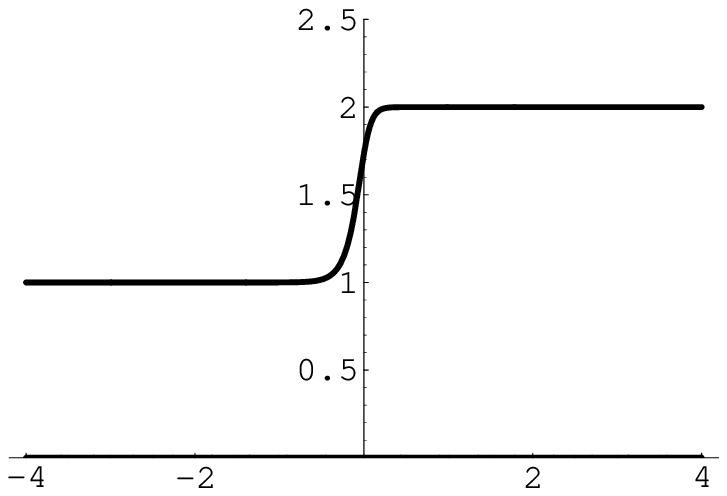}\hspace{1cm}
\includegraphics[height=3.cm]{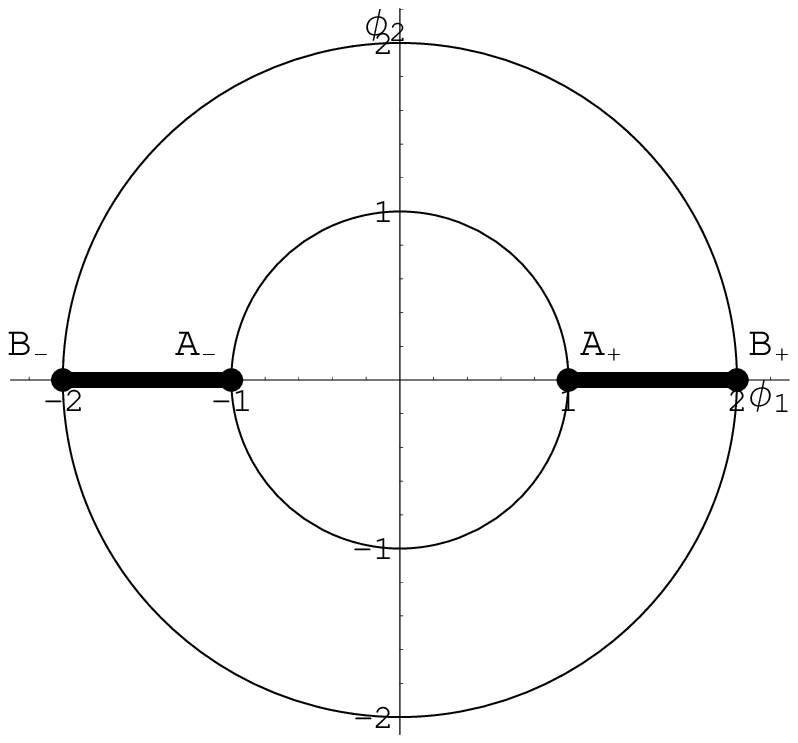}\hspace{1cm}
\includegraphics[height=3.cm]{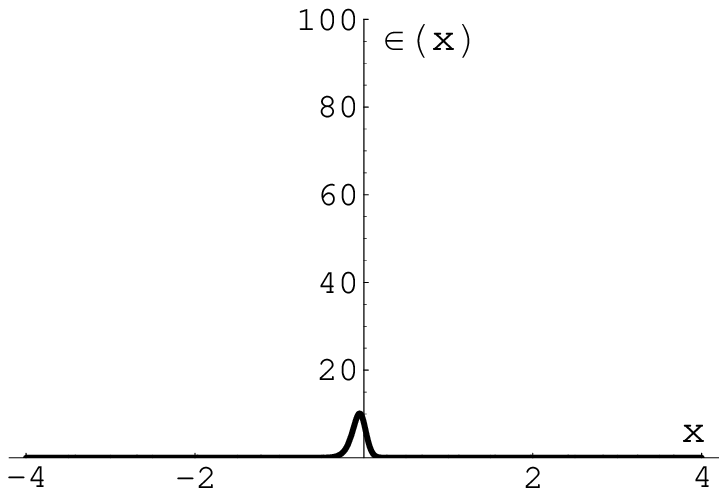}}
\caption{\small \textit{Solitary waves ${\rm K}_1^{AB}$: a) Form
Factor b) Orbit and c) Energy Density.}}
\end{figure}
\vspace{0.1cm}

$\bullet$ $K2_1^{AA^*}$: The orbit of the second kind of solitary
waves on the $\phi_2$ axis is the $[A_-,A_+]$ interval. The kink
profile of these kinks connects these points in the vacuum orbit and
is also easy to determine:
\[
\vec{\phi}^{{\rm K2}_1^{AA^*}}(x)= 2 \cos \left[ \frac{2}{3} \arctan
e^{6\sqrt{2} \bar{x}}+\frac{4\pi}{3} \right] \vec{e}_1 \qquad .
\]
Despite being analytically different, the behavior of these kinks
is completely analogous to that of the bizarre $K2_1^{AA^*}$
solitary waves of the A1 model. The energy density and total
energy of these kinks are:
\[
{\cal E}^{{\rm K2}_1^{AA^*}}(x) = 32 \, {\rm sech}^2 6 \sqrt{2}
\bar{x} \, \sin^2 \left( \frac{2}{3} \arctan e^{6 \sqrt{2}
\bar{x}}+{4\pi\over 3} \right)
\]
\[
E[{\rm
K2}_1^{AA^*}]=\left|W(A_-,\phi_2=0)-\lim_{\varepsilon\rightarrow
0^-}W(\varepsilon,\phi_2=0)\right|+\left|\lim_{\varepsilon\rightarrow
0^+}W(\varepsilon,\phi_2=0)-W(A_+,\phi_2=0)\right|=\frac{76
\sqrt{2}}{15} \qquad ,
\]
see Figure 3 to find qualitative plots of their properties.

\subsection{Solitary waves in the bulk of the moduli space}

The Hamilton-Jacobi orbit (\ref{eq:orbgen}) and time schedule
(\ref{eq:espgen}) are in this case the quadratures:
\begin{eqnarray*}
(-1)^\alpha \int \frac{dR}{R^2(1-R^2)(4-R^2)}-(-1)^\beta \int
\frac{d \varphi}{\sigma \sin \varphi}&=&\sqrt{2} \gamma_1 \\
\int \frac{dR}{(1-R^2)(4-R^2)}&=&\sqrt{2} \bar{x} \qquad .
\end{eqnarray*}
Integration of these equations provides the analytic expressions
for kink orbits and form factors:
\begin{eqnarray}
(-1)^\alpha \left[\frac{-1}{4R}+\frac{1}{6} \log
\frac{(R-2)^\frac{1}{8} (R+1)}{(R+2)^\frac{1}{8} (R-1)} \right] -
\frac{(-1)^\beta}{\sigma} \log \tan \frac{\varphi}{2} \label{eq:kinoB1}&=&\sqrt{2} \gamma_1 \\
\frac{1}{12} \log \frac{(R-2) (R+1)^2}{(R+2)(R-1)^2} &=& \sqrt{2}
\bar{x} \label{eq:kinkfB1} \qquad .
\end{eqnarray}

Solitary wave solutions only arise from equations
(\ref{eq:kinoB1}) and (\ref{eq:kinkfB1}) if $1\leq R \leq 2$ or
$0\leq R \leq 1$. Let us denote $\Theta=\arctan e^{6\sqrt{2}
\bar{x}}+{2\pi\over 3}\aleph$ in  the first range, $\Theta=\arctan
e^{6\sqrt{2} \bar{x}}+{4\pi\over 3}$ in the second range, and let
us define the functions:
\begin{equation}
\Omega_1(x)=\frac{\cos \frac{2\Theta}{3}+\frac{1}{2}}{\cos
\frac{2\Theta}{3}-\frac{1}{2}} \hspace{0.7cm};\hspace{0.7cm}
\Omega_2(x)=\frac{\Omega_1(x)-1}{\Omega_1(x)+1}=\frac{1}{2} \sec
\frac{2\Theta}{3}\hspace{0.7cm};\hspace{0.7cm}\Lambda (x)=\tan
\frac{\Theta}{3} \label{eq:notafun} \qquad .
\end{equation}

$\bullet$ ${\rm K}_2^{AB}(\gamma_1)$: If $1\leq R \leq 2$, the
solutions in Cartesian coordinates are:
\[
\vec{\phi}^{{\rm K}_2^{AB}(\gamma_1)}(x)= \phi_1^{{\rm
K}_2^{AB}(\gamma_1)}(x)\, \vec{e}_1+\phi_2^{{\rm
K}_2^{AB}(\gamma_1)}(x)\, \vec{e}_2
\]
\begin{eqnarray*}
\vec{\phi}^{{\rm
K}_2^{AB}(\gamma_1)}(x)&=&\frac{1}{\Omega_2(\bar{x})} \left[
\frac{2}{1+e^{-2\sqrt{2}\sigma \gamma_1}\, e^{-\frac{\sigma}{2}
\Omega_2(\bar{x})} \, \Omega_1^\frac{\sigma}{3}(\bar{x})\,
\Lambda^\frac{\sigma}{12}(\bar{x})}-1\right]\vec{e}_1 \\&+&
(-1)^\alpha \, \frac{2 \, e^{\sqrt{2}\sigma \gamma_1}\,
e^{\frac{\sigma}{4}\Omega_2(\bar{x})}\,
\Omega_1^\frac{\sigma}{6}(\bar{x})\,
\Lambda^\frac{\sigma}{24}(\bar{x}) }{\Omega_2(\bar{x})\left(
e^{2\sqrt{2}\sigma \gamma_1}\,
e^{\frac{\sigma}{2}\Omega_2(\bar{x})}+\Omega_1^\frac{\sigma}{3}(\bar{x})\,
\Lambda^\frac{\sigma}{12}(\bar{x}) \right)}\vec{e}_2 \qquad , \qquad
\alpha=0,1 \qquad .
\end{eqnarray*}
The kink orbits belonging to these two one-parametric families
connect either $A_-$ to $B_+$ if $\aleph=0$, or $A_+$ to $B_-$ if
$\aleph=0$, and are confined inside the annulus bounded by the
circles $C_1$ and $C_2$, see Figure 10. The energy density of each
member of these kink families is localized around two distinct
points, see Figure 10. For sufficiently negative, values of
$\gamma_1$ the two lumps are composed of one $K1_1^{AA^*}$ and one
$K_1^{AB}$ basic kinks. In the opposite regime, for sufficiently
positive values of $\gamma_1$, two lumps arise again, close to one
$K_1^{BB^*}$ and one $K_1^{AB}$ basic kinks. Note that the basic
kinks appear at the boundary of the moduli space when either
$\gamma_1=-\infty$ or $\gamma_1=\infty$. Intermediate tuning of
$\gamma_1$ continuously shifts from one configuration to the
other, see Figure 11. These features are synthesized in the
notation ${\rm K}_2^{AB}(\gamma_1)$.

\begin{figure}[htb]
\centerline{\includegraphics[height=3.cm]{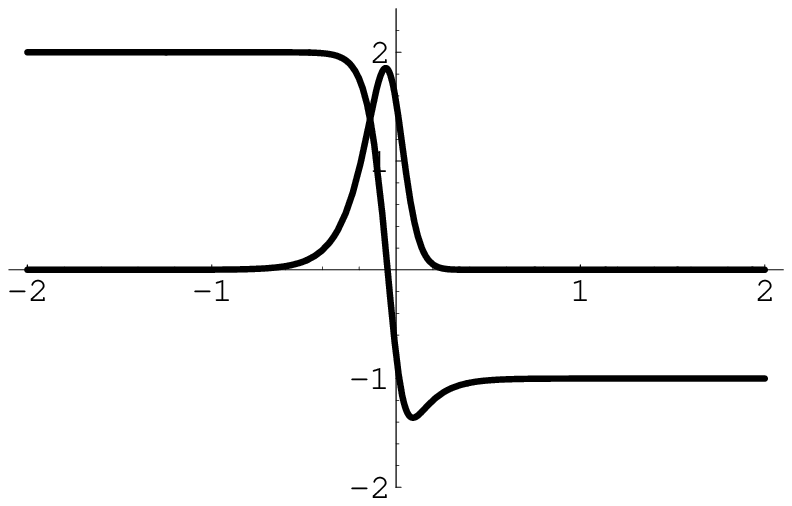}\hspace{1cm}
\includegraphics[height=3.cm]{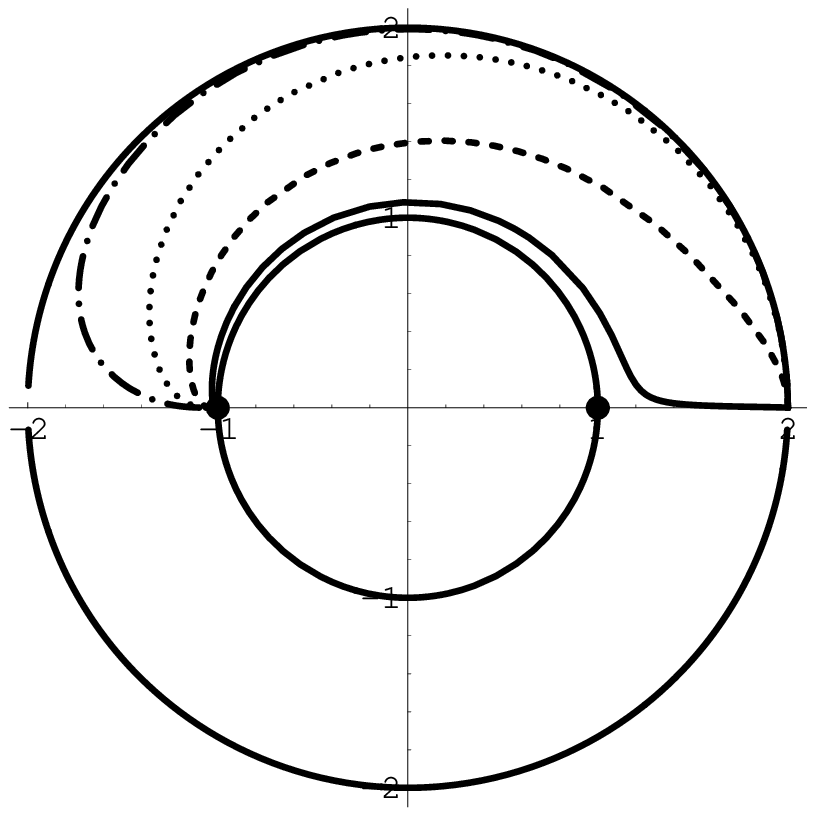}}
\caption{\small \textit{Solitary wave family ${\rm
K}_2^{AB}(\gamma_1)$: a) form factor and b) Orbits for values
$\gamma_1=0,\pm 1,\pm 2$.}}
\end{figure}

\begin{figure}[htb]
\centerline{\includegraphics[height=2.3cm]{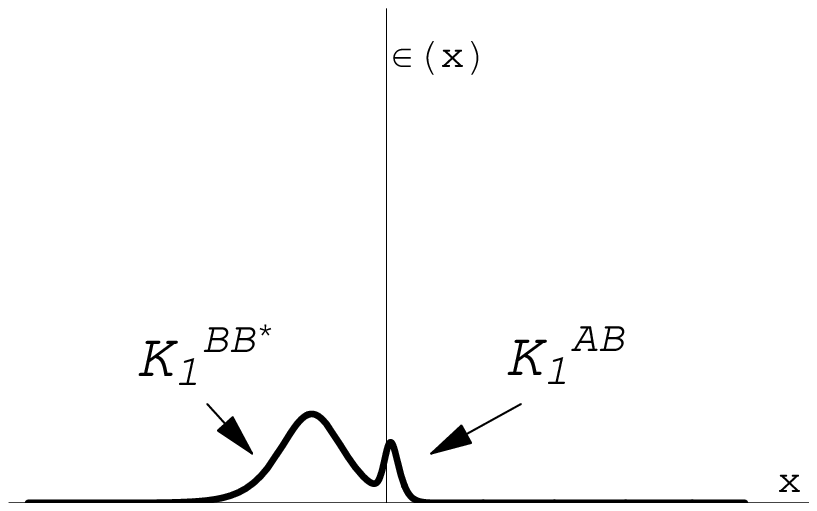}
\includegraphics[height=2.3cm]{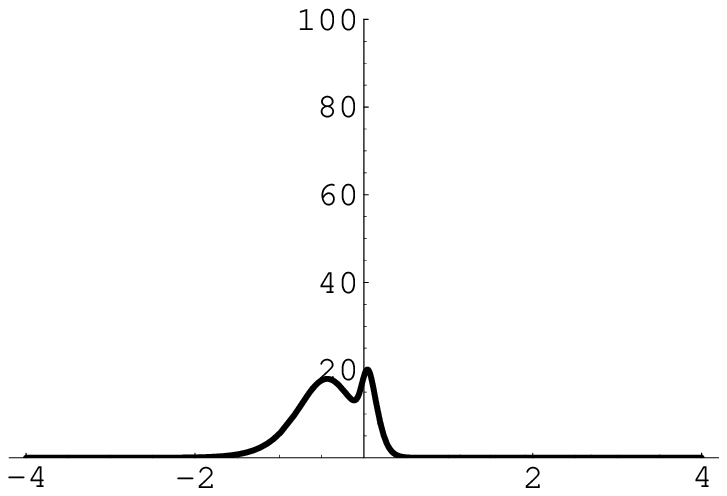}
\includegraphics[height=2.3cm]{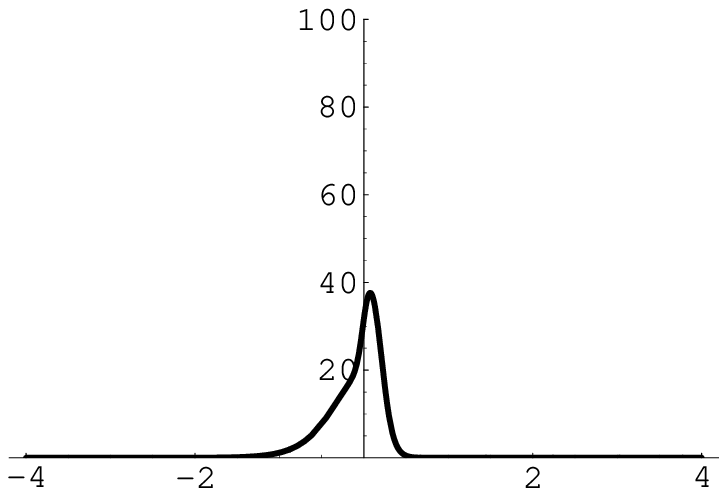}
\includegraphics[height=2.3cm]{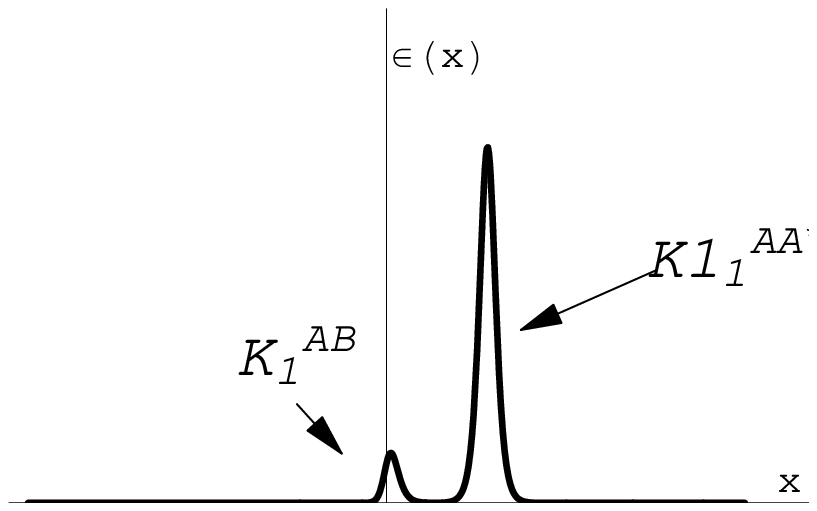}}
\caption{\small \textit{Energy Density of the solitary wave family
${\rm K}_2^{AB}(\gamma_1)$ for increasing values of $\gamma_1$.}}
\end{figure}

\vspace{0.1cm}

$\bullet$ ${\rm K}_3^{AA}(\gamma_1^+,\gamma_1^-)$: Inside the
$C_1$, $0<R<1$, there are kink solutions completely analogous to
the ${\rm K}_3^{AA}(\gamma_1^+,\gamma_1^-)$ solitary waves of the
$A1$ model, see Figure 6. The kink profiles have the form:
\[
\vec{\phi}^{{\rm K}_3^{AA^*}(\gamma_1^+,\gamma_1^-)}(x)=\left\{
\begin{array}{l} \phi_1^{K_3^{AO}}(\bar{x};\gamma_1^-)\,
\vec{e}_1+\phi_2^{K_3^{AO}}(\bar{x};\gamma_1^-) \, \vec{e}_2 \mbox{ if } \bar{x} \leq 0\\
\phi_1^{K_3^{AO}}(\bar{x};\gamma_1^+)\,
\vec{e}_1-\phi_2^{K_3^{AO}}(\bar{x};\gamma_1^+) \, \vec{e}_2
\mbox{ if } \bar{x}
> 0 \end{array} \right. \qquad ,
\]
where
\begin{eqnarray*}
\vec{\phi}^{{\rm
K}_2^{AO}(\gamma_1)}(x)&=&\frac{1}{\Omega_2(\bar{x})} \left[
\frac{2}{1+e^{-2\sqrt{2}\sigma \gamma_1}\, e^{-\frac{\sigma}{2}
\Omega_2(\bar{x})} \, \Omega_1^\frac{\sigma}{3}(\bar{x})\,
\Lambda^\frac{\sigma}{12}(\bar{x})}-1\right]\vec{e}_1 \\&+&
(-1)^\alpha \, \frac{2 \, e^{\sqrt{2}\sigma \gamma_1}\,
e^{\frac{\sigma}{4}\Omega_2(\bar{x})}\,
\Omega_1^\frac{\sigma}{6}(\bar{x})\,
\Lambda^\frac{\sigma}{24}(\bar{x}) }{\Omega_2(\bar{x})\left(
e^{2\sqrt{2}\sigma \gamma_1}\,
e^{\frac{\sigma}{2}\Omega_2(\bar{x})}+\Omega_1^\frac{\sigma}{3}(\bar{x})\,
\Lambda^\frac{\sigma}{12}(\bar{x}) \right)}\vec{e}_2 \qquad ,
\qquad \alpha=0,1 \qquad ,
\end{eqnarray*}
and $\Theta=\arctan e^{6\sqrt{2} \bar{x}}+{4\pi\over 3}$ in the
$0<R<1$ range.

In particular , the limiting case $K_3^{AA}(\gamma_1^+,-\infty)$
gives a solitary wave identical to the $K_2^{AA}(\gamma_1)$ kink
of the A1 model. To obtain the kink profile one replaces
$\phi_1^{K_3^{AO}}(\bar{x};\gamma_1^-)\,
\vec{e}_1+\phi_2^{K_3^{AO}}(\bar{x};\gamma_1^-) \, \vec{e}_2$ by
$\vec{\phi}^{K2_1^{AA^*}}(\gamma_1)(x)=2{\rm cos}\Theta \vec{e}_1$
on the left ($\bar{x}<0$). We recall that these solutions connect
a point in $A$ with itself through a trajectory that crosses the
origin, see Figure 4.

In sum,  the structure of the solitary wave variety in the B1
model is as follows: There exist four kinds of basic lumps
associated with the ${\rm K1}_1^{AA^*}$, ${\rm K2}_1^{AA^*}$,
${\rm K}_1^{BB}$ and ${\rm K}_1^{AB}$ kinks, which display
different distributions of one-point localized energy densities.
The ${\rm K}_3^{AA}(\gamma_1^+,\gamma_1^-))$, ${\rm
K}_2^{AA}(\gamma_1)$, and ${\rm K}_2^{AB}(\gamma_1)$ kinks,
however, are composite solitary waves.  The ${\rm
K}_3^{AA}(\gamma_1^+,\gamma_1^-)$ kinks are combinations of two
${\rm K1}_1^{AA^*}$ and one ${\rm K2}_1^{AA^*}$ basic lumps; the
${\rm K}_2^{AA}(\gamma_1)$ are formed by one ${\rm K1}_1^{AA^*}$
and one ${\rm K2}_1^{AA^*}$ basic lumps, the ${\rm
K}_2^{AB}(\gamma_1)$ kinks exhibit an orbit dependent structure
that varies between the limiting combination formed by either the
${\rm K1}_1^{AA^*}$ and ${\rm K}_1^{AB}$ kinks or the ${\rm
K1}_1^{BB^*}$ and ${\rm K}_1^{AB}$ kinks. Therefore, the following
kink energy sum rules hold:
\[
E[K_3^{AA}]=2E[K1_1^{AA}]+E[K2_1^{AA}]=4\sqrt{2}\sigma+{76\over
15}\sqrt{2}
\]
\[
E[K_2^{AA}]=E[K1_1^{AA}]+E[K2_1^{AA}]=2\sqrt{2}\sigma+{76\over
15}\sqrt{2}
\]
\[
E[K_2^{AB}]=E[K_1^{AB}]+E[K_1^{BB}]=E[K_1^{AB}]+E[K1_1^{AA}]=2\sqrt{2}\sigma+{22\over
15}\sqrt{2} \qquad .
\]
With respect to stability, all the kinks inside the $C_1$ circle
behave like their cousins in the A1 model. All the kinks in the
$1\leq R\leq2$ annulus are stable.

\section{The A2 model}

In this Section we analyze the deformation induced on the potential
$v^{(A)}(\phi_1,\phi_2)$ (\ref{eq:va}) by $w^{(2)}(\phi_1,\phi_2)$
(\ref{eq:w2}). The action functional (\ref{eq:action}) with
potential energy
\begin{equation}
U_{\rm A2}(\phi_1,\phi_2)=(\phi_1^2+\phi_2^2-1)^2+
\frac{\sigma^2}{2(\phi_1^2+\phi_2^2)}\left(1-
 \frac{\phi_1}{\sqrt{\phi_1^2+\phi_2^2}}
 \right) \label{eq:pota2}
\end{equation}
governs the dynamics. This deformation explicitly breaks the
$SO(2)$ symmetry to the discrete group $\mathbb{Z}_2$ generated by
the reflection $\phi_2\rightarrow -\phi_2$. The set ${\cal M}$ of
zeroes of $U(\phi_1,\phi_2)$ has only one element:
\[
{\cal M}=\{ \, A= (1,0) \,  \} \qquad \qquad .
\]
In this model, the second order partial differential equations
\begin{eqnarray*}
\frac{\partial^2 \phi_1}{\partial t^2}-\frac{\partial^2
\phi_1}{\partial x^2} &=& 4 \phi_1
(1-\phi_1^2-\phi_2^2)+\frac{\sigma^2}{2}
\frac{\phi_2^2-2\phi_1^2+2\phi_1
\sqrt{\phi_1^2+\phi_2^2}}{(\phi_1^2+\phi_2^2)^\frac{5}{2}} \\
\frac{\partial^2 \phi_1}{\partial t^2}-\frac{\partial^2
\phi_1}{\partial x^2} &=& 4 \phi_2
(1-\phi_1^2-\phi_2^2)+\frac{\sigma^2}{2}\left[\frac{2\phi_2}{(\phi_1^2+
\phi_2^2)^2}-\frac{3\phi_1\phi_2}{(\phi_1^2+\phi_2^2)^\frac{5}{2}}
\right]
\end{eqnarray*}
are the field equations. The asymptotic conditions
(\ref{eq:asymtotic}) guaranteeing finite energy compel the
solitary wave solutions to link the only minimum $A$ with itself.
The general results in Section \S. 2 ensure that the $C_1\equiv
\phi_1^2+\phi_2^2=1$ circle and the $r_1\equiv \, \phi_2=0 \, \cup
\, \phi_1>0$ abscissa half axis are separatrix curves. Unlike the
A1 and B1 models (see Section \S. 3 and \S. 4) the singularity at
the origin is of different nature in this case. We know from
previous Sections that there exists a path $\phi_2=0$ along which
the singularity is not felt in either model A1 or B1. The
potential energy $U^{\rm A2}(\phi_1,\phi_2)$, however, reduces on
the abscissa axis to:
\[
U^{\rm A2}(\phi_1,0)=(\phi_1^2-1)^2+ \frac{\sigma^2}{2 \phi_1^2}
\left(1-\,{\rm sign}\,(\phi_1) \right)
\]
and the singularity is always felt in the $\phi_1<0$ negative
half-axis. There is no escape to kink orbits that enter through
the positive abscissa half-axis, and finite energy solitary waves
are pushed away from the origin.

Nevertheless, there are kink orbits on the $C_1\equiv
\phi_1^2+\phi_2^2=1$ circle.

\vspace{0.1cm}

$\bullet$ ${\rm K}_1^{AA}$: Plugging this curve into
(\ref{eq:edobogo}) we obtain
\[
\vec{\phi}^{{\rm K}_1^{AA}}(x)= \left(2 \tanh^2 \frac{\sigma
\bar{x}}{\sqrt{2}} -1 \right)\vec{e}_1 +2 \, {\rm sech}\,
\frac{\sigma \bar{x}}{\sqrt{2}} \tanh \frac{\sigma
\bar{x}}{\sqrt{2}} \,\, \vec{e}_2 \qquad ,
\]
a solitary wave connecting the point $A$ with itself when $x$ varies
from $-\infty$ to $\infty$, see Figure 12(a,b). The energy density
\[
{\cal E}^{{\rm K}_1^{AA}}(x)=2\,\sigma^2 \, {\rm sech}^2\,
\frac{\sigma \bar{x}}{\sqrt{2}} \hspace{2cm} , \hspace{2cm} E[{\rm
K}_1^{AA}]=4\sqrt{2} \sigma
\]
is localized at one point so that the ${\rm K}_1^{AA}$ is a basic
lump, see Figure 12(c).

\begin{figure}[htb]
\centerline{\includegraphics[height=3.cm]{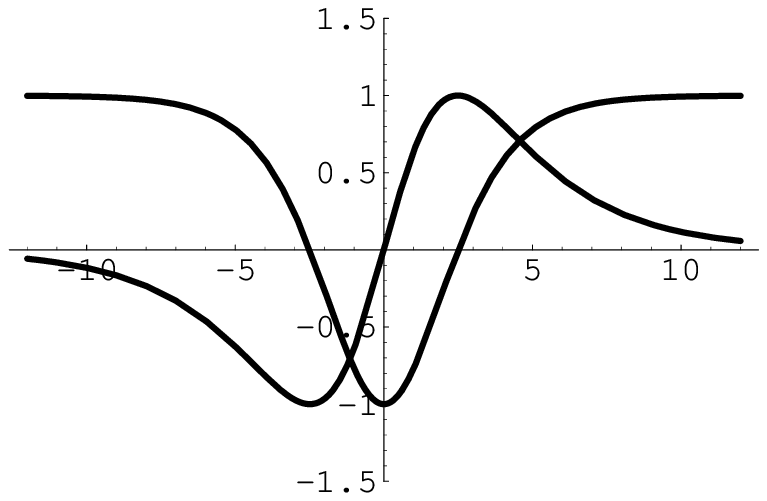}\hspace{1cm}
\includegraphics[height=3.cm]{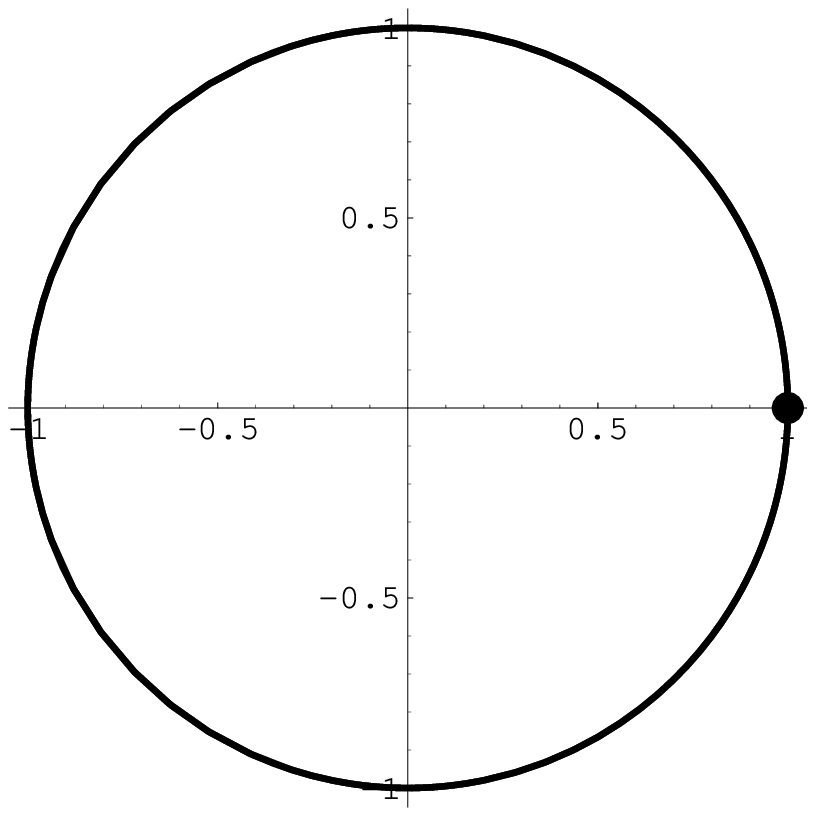}\hspace{1cm}
\includegraphics[height=3.cm]{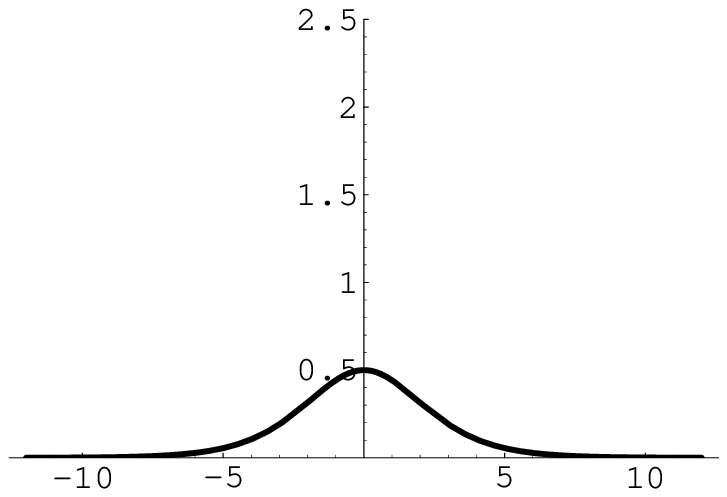}}
\caption{\small \textit{Solitary waves ${\rm K}_1^{AA}$: a) Factor
Form, b) Orbit and c) Energy Density.}}
\end{figure}

\vspace{0.1cm}

 Again we are dealing with a mechanical analogous system of Liouville Type II such that:
\[
 f(R)=(R^2-1)^2 \hspace{2cm} , \hspace{2cm}
g(\varphi)=\sigma^2 \sin^2 \frac{\varphi}{2} \qquad .
\]
In Cartesian coordinates the superpotentials read:
\[
\left\{\begin{array}{c}
W^+(\phi_1,\phi_2)=(-1)^\alpha\sqrt{2}\sqrt{\phi_1^2+\phi_2^2}\left(\frac{\phi_1^2+\phi_2^2}{3}-1\right)
-2\sigma(-1)^\beta\sqrt{1+\sqrt{\frac{\phi_1^2}{\phi_1^2+\phi_2^2}}}
\qquad , \qquad \phi_2>0 \\
W^-(\phi_1,\phi_2)=(-1)^\alpha\sqrt{2}\sqrt{\phi_1^2+\phi_2^2}\left(\frac{\phi_1^2+\phi_2^2}{3}-1\right)
+2\sigma(-1)^\beta\sqrt{1+\sqrt{\frac{\phi_1^2}{\phi_1^2+\phi_2^2}}}
\qquad , \qquad \phi_2<0 \end{array}\right.
\]
and we realize that the kink energy is a topological bound
\[
E[{\rm K}_1^{AA}]=\left|W^+(A)-W^-(A)\right|=4\sqrt{2} \sigma
\]
giving the winding number of the kink orbit around the origin. This
fact ensures stability for the ${\rm K}_1^{AA}$ kink.

The Hamilton-Jacobi theory yields the remaining solutions confined
inside the unit circle $C_1$. In this model, the Hamilton-Jacobi
formulae (\ref{eq:orbgen}) and (\ref{eq:espgen}) or, equivalently,
the Bogomolny first-order ODE system (\ref{eq:edobogo}) become:
\begin{eqnarray}
(-1)^\alpha \int \, \frac{dR}{R^2\,(R^2-1)}-(-1)^\beta \int \,
\frac{d\varphi}{{2}\sqrt\sigma\,{\rm sin}\,{\varphi\over
2}}&=&\sqrt{2}\gamma_1 \label{eq:IIA2or1}\\\int \,
\frac{dR}{R^2-1}&=&\sqrt{2}\bar{x} \qquad .\label{eq:IIA2or2}
\end{eqnarray}
Integration of (\ref{eq:IIA2or1}) gives the kink orbits:
\begin{equation}
\log
\left(\left|\frac{R^K(x)+1}{R^K(x)-1}\right|^{\frac{(-1)^\alpha}{2}}\cdot
e^{-{1\over R^K(x)}}\right)-\log\left|{\rm
tan}\frac{\varphi^K(x)}{4}\right|^{\frac{(-1)^\beta}{\sigma}}=\sqrt{2}\gamma_1
\qquad .\label{eq:polo}
\end{equation}
The kink form factor (kink profile) in the $R$-variable is
obtained by integrating (\ref{eq:IIA2or2}):
\begin{equation}
R^K(x)={\rm tanh}\left(\sqrt{2}\left|\bar{x}
\right|\right)\label{eq:kppol}\qquad .
\end{equation}\vspace{0.1cm}
Plugging this solution into (\ref{eq:IIA2or1}) the kink profile
for the angular variable is found:
\begin{equation}
\varphi^K(x)=4{\rm arctan}\left[{\rm
exp}\left(\sqrt{2}\sigma(\bar{x}+\gamma_1)-\sigma{\rm
cotan}\sqrt{2}\bar{x}\right)\right] \label{eq:kpol2} \qquad .
\end{equation}
Back in Cartesian coordinates, we obtain:
\begin{eqnarray*}
\vec{\phi}(x)&=& \tanh \sqrt{2} |\bar{x}| \left\{1-2\,{\rm sech}\,
\left[\frac{\sigma}{2}\left( \gamma_1-\sqrt{2} \bar{x} + {\rm
cotanh}\, \sqrt{2}\bar{x} \right) \right] \right\}\vec{e}_1 + \\
&+& 2 \tanh \sqrt{2} |\bar{x}| \tanh \left( \gamma_1-\sqrt{2}
\bar{x} + {\rm cotanh}\, \sqrt{2}\bar{x} \right) {\rm sech}\left(
\gamma_1-\sqrt{2} \bar{x} + {\rm cotanh}\, \sqrt{2}\bar{x}
\right)\vec{e}_2 \qquad .
\end{eqnarray*}
Some orbits for these solutions are depicted in Figure 13. Here,
we notice that all solutions of this kind head towards the origin;
thus, they cannot be regarded as finite-energy solutions.

\begin{figure}[htb]
\centerline{
\includegraphics[height=3.cm]{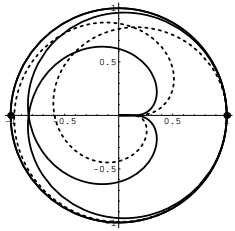}}
\caption{\small \textit{Trajectories of solutions confined in the
unit circle $C_1$. }}
\end{figure}

In sum, in model $A2$ there exists only one basic solitary wave,
the ${\rm K}_1^{AA}$ kink, and no composite solitary waves arise.

\section{The B2 Model}

Finally, in this last section we study  the deformation $w^{(2)}$
(\ref{eq:w2}) induced on $v^{(B)}$ (\ref{eq:vb}). Therefore, the
potential term
\begin{equation}
U_{\rm
B2}(\phi_1,\phi_2)=(\phi_1^2+\phi_2^2-1)^2(\phi_1^2+\phi_2^2-a^2)^2+
\frac{\sigma^2}{2(\phi_1^2+\phi_2^2)}\left(1-
 \frac{\phi_1}{\sqrt{\phi_1^2+\phi_2^2}}
 \right) \label{eq:potb2}
\end{equation}
enters the functional action (\ref{eq:action}). We set the $a=2$
value bearing in mind that models of B2 type characterized by
other non-null values of the parameter $a$ show a very similar
behavior. As in the $A2$ model, the second summand on the right
hand side of (\ref{eq:potb2}) explicitly breaks the $SO(2)$
symmetry of the unperturbed system (\ref{eq:vb}) to the discrete
group $\mathbb{Z}_2$ generated by the $\phi_2\rightarrow -\phi_2$
reflection. The non-deformed manifold of zeroes ${\cal
M}_{\sigma=0}=S^1_{r=1}\cup S^1_{r=2}$, the disjoint union of two
circles, becomes the discrete set of two elements:
\[
{\cal M}=\{ \, A= (1,0) \, \, ; \, \,  B=(2,0)  \} \qquad \qquad
.
\]
The Euler-Lagrange equations form the system of second-order PDEs:
\begin{eqnarray*}
\frac{\partial^2 \phi_1}{\partial t^2}-\frac{\partial^2
\phi_1}{\partial x^2} &=& 4 \phi_1
(1-\phi_1^2-\phi_2^2)(4-\phi_1^2-\phi_2^2)(5-2\phi_2^2-2\phi_2^2)+
\frac{\sigma^2}{2}
\frac{\phi_2^2-2\phi_1^2+2\phi_1
\sqrt{\phi_1^2+\phi_2^2}}{(\phi_1^2+\phi_2^2)^\frac{5}{2}} \\
\frac{\partial^2 \phi_1}{\partial t^2}-\frac{\partial^2
\phi_1}{\partial x^2} &=& 4 \phi_2
(1-\phi_1^2-\phi_2^2)(4-\phi_1^2-\phi_2^2)(5-2\phi_2^2-2\phi_2^2)+\frac{\sigma^2}{2}\left[\frac{2\phi_2}{(\phi_1^2+
\phi_2^2)^2}-\frac{3\phi_1\phi_2}{(\phi_1^2+\phi_2^2)^\frac{5}{2}}
\right] \qquad .
\end{eqnarray*}
Solitary waves are solutions of these equations complying with the
asymptotic conditions (\ref{eq:asymtotic}). Thus, the kink must
connect a point, either $A$ or $B$, with itself or must link $A$
with $B$. The general results in  section \S. 2 ensure that the
separatrix curves in this model will be the $C_1\equiv
\phi_1^2+\phi_2^2=1$, $C_2\equiv \phi_1^2+\phi_2^2=4$ circles, and
the $r_1\equiv \, \phi_2=0 \, \cup \, \phi_1>0$ abscissa
half-axis. The nature of the singularity arising in the
(\ref{eq:potb2}) potential is identical to the singularity of the
$A2$ model, see section \S. 5. Each solution that crosses through
the origin cannot be regarded as having finite energy.

The analogous mechanical system is a Liouville Type II system such
that:
\[
f(R)=(R^2-1)^2(R^2-4)^2 \hspace{2cm} , \hspace{2cm}
g(\varphi)=\sigma^2 \sin^2 \frac{\varphi}{2} \qquad .
\]
Accordingly, in Cartesian coordinates the superpotentials read:
\[
\begin{array}{c} W^+(\phi_1,\phi_2)=\sqrt{2}
(-1)^\alpha\sqrt{\phi_1^2+\phi_2^2}\left[ \frac{1}{5}
(\phi_1^2+\phi_2^2)^2-\frac{5}{3}(\phi_1^2+\phi_2^2)+4 \right]
-2\sigma(-1)^\beta\sqrt{1+\sqrt{\frac{\phi_1^2}{\phi_1^2+\phi_2^2}}}
\quad , \quad \phi_2>0 \\
W^-(\phi_1,\phi_2)=\sqrt{2}
(-1)^\alpha\sqrt{\phi_1^2+\phi_2^2}\left[ \frac{1}{5}
(\phi_1^2+\phi_2^2)^2-\frac{5}{3}(\phi_1^2+\phi_2^2)+4 \right]
+2\sigma(-1)^\beta\sqrt{1+\sqrt{\frac{\phi_1^2}{\phi_1^2+\phi_2^2}}}
\quad , \quad \phi_2<0 \end{array}
\]

Bearing this in mind we now describe the solitary wave or kink
variety of the model:

\vspace{0.1cm}
\subsection{Solitary waves at the boundary of the moduli space}

$\bullet$ ${\rm K}_1^{AA}$: Proposition 1 in section \S. 2
guarantees the existence of solitary waves with orbits running on
the unit circle $C_1\equiv \phi_1^2+\phi_2^2=1$. Integration of
the first-order equations (\ref{eq:edobogo}) for configurations
living on $C_1$ provides the solitary wave solutions:
\[
\vec{\phi}^{{\rm K}_1^{AA}}(x)= \left(2 \tanh^2 \frac{\sigma
\bar{x}}{\sqrt{2}} -1 \right)\vec{e}_1 +2 \, {\rm sech}\,
\frac{\sigma \bar{x}}{\sqrt{2}} \tanh \frac{\sigma
\bar{x}}{\sqrt{2}} \,\, \vec{e}_2 \qquad .
\]
These solitary waves are non-topological -but stable- kinks
identical to the $K_1^{AA}$ kinks of the $A2$ model, encircling
the origin in a orbit that starts and ends at $A$, see Figure
14(a,b). They are of course basic lumps because their energy
density
\[
{\cal E}^{K_1^{AA}}(x)=2\,\sigma^2 \, {\rm sech}^2\, \frac{\sigma
\bar{x}}{\sqrt{2}} \hspace{2cm} , \hspace{2cm} E[{\rm
K}_1^{AA}]=\left|W^+(A)-W^-(A)\right|=4\sqrt{2} \sigma
\]
is localized around a point, see Figure 14(c), and their total
energy is a topological bound proportional to the winding number of
the kink orbit around the origin.

\begin{figure}[htb]
\centerline{\includegraphics[height=3.cm]{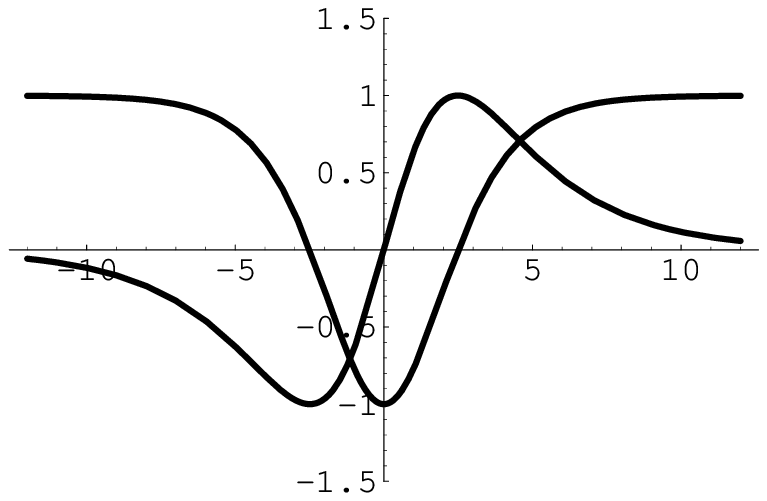}\hspace{1cm}
\includegraphics[height=3.cm]{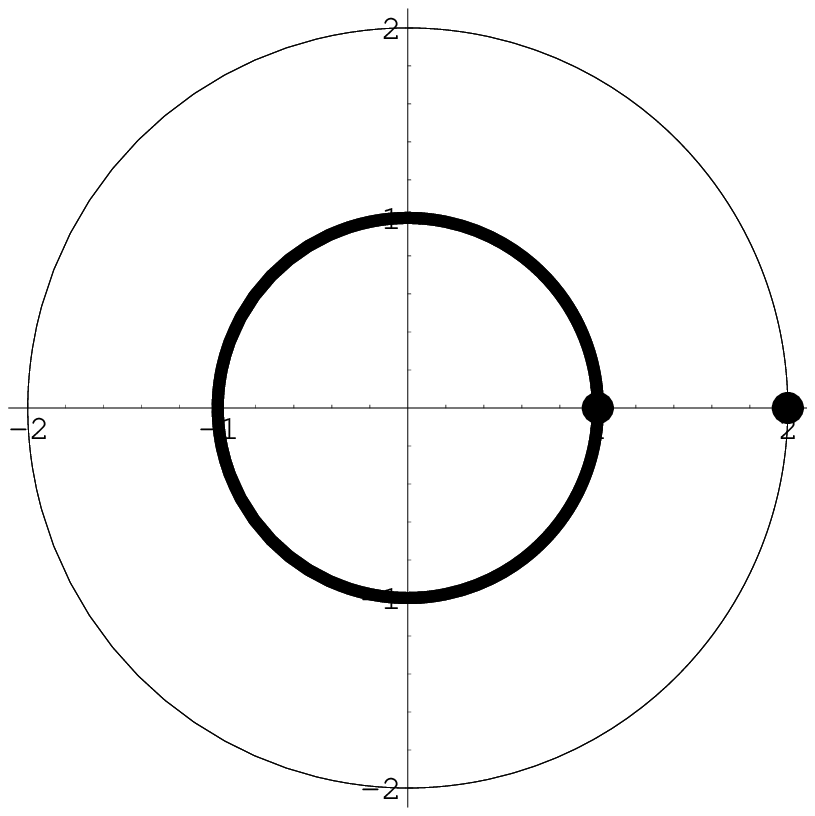}\hspace{1cm}
\includegraphics[height=3.cm]{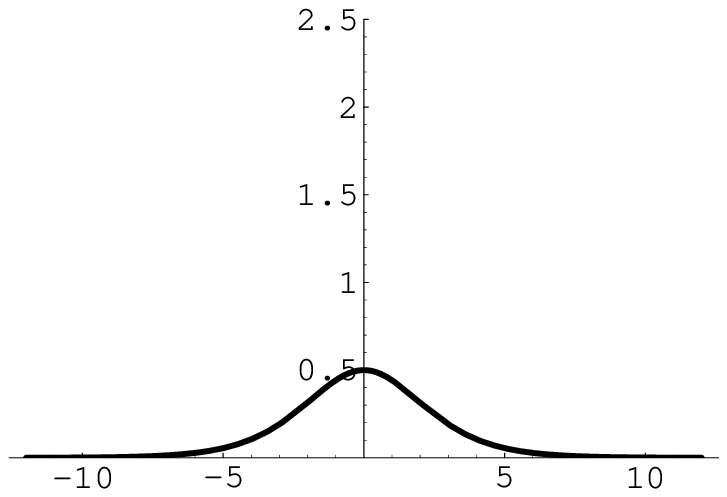}}
\caption{\small \textit{Solitary waves ${\rm K}_1^{AA}$: a) Form
factor, b) Orbit, and c) Energy Density.}}
\end{figure}

\vspace{0.1cm}

$\bullet$ ${\rm K}_1^{BB}$: By the same token there are solitary
waves whose orbit is the $C_2\equiv \phi_1^2+\phi_2^2=4$ circle.
Plugging this trial orbit into the first-order equations we obtian
via one quadrature the solitary wave solutions:
\[
\vec{\phi}^{{\rm K}_1^{BB}}(x)= \left(4 \tanh^2 \frac{\sigma
\bar{x}}{4 \sqrt{2}} -2 \right)\vec{e}_1 +4 \, {\rm sech}\,
\frac{\sigma \bar{x}}{4\sqrt{2}} \tanh \frac{\sigma \bar{x}}{4
\sqrt{2}} \,\, \vec{e}_2 \qquad .
\]
These non-topological kinks have orbits departing from and arriving
at the same point $B$, see Figure 15(a,b). Their energy density
\[
{\cal E}^{K_1^{BB}}(x)=\,\frac{\sigma^2}{2} \, {\rm sech}^2\,
\frac{\sigma \bar{x}}{4 \sqrt{2}} \hspace{2cm} , \hspace{2cm} E[{\rm
K}_1^{BB}]=\left|W^+(B)-W^-(B)\right|=4\sqrt{2} \sigma
\]
is again localized at a point, but the energy distribution is
different from the energy distribution of the $K_1^{AA}$ kinks.
Thus, the $K_1^{BB}$ kinks, besides belonging to a different
topological sector of the configuration space, are basic lumps or
particles different from $K_1^{AA}$ kinks, see Figure 15(c).
Moreover, $K_1^{BB}$ kinks are also stable because their energy is
given by the winding number of the kink orbit around the origin.

\begin{figure}[htb]
\centerline{\includegraphics[height=3.cm]{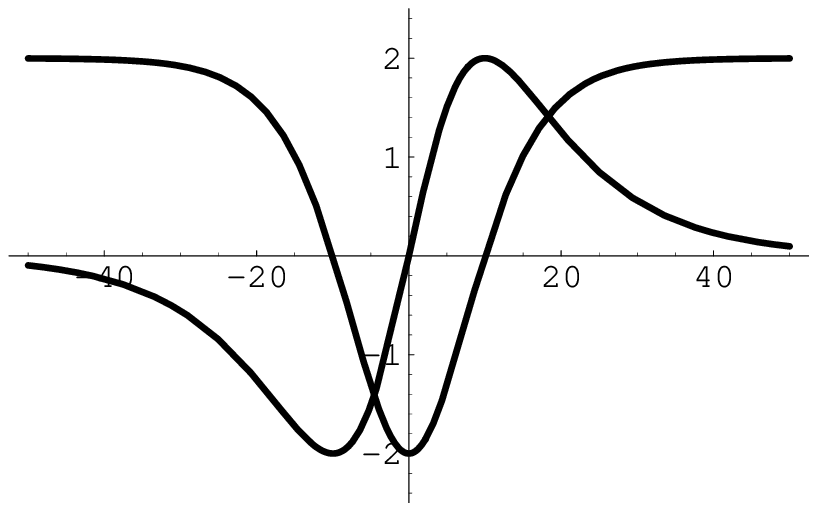}\hspace{1cm}
\includegraphics[height=3.cm]{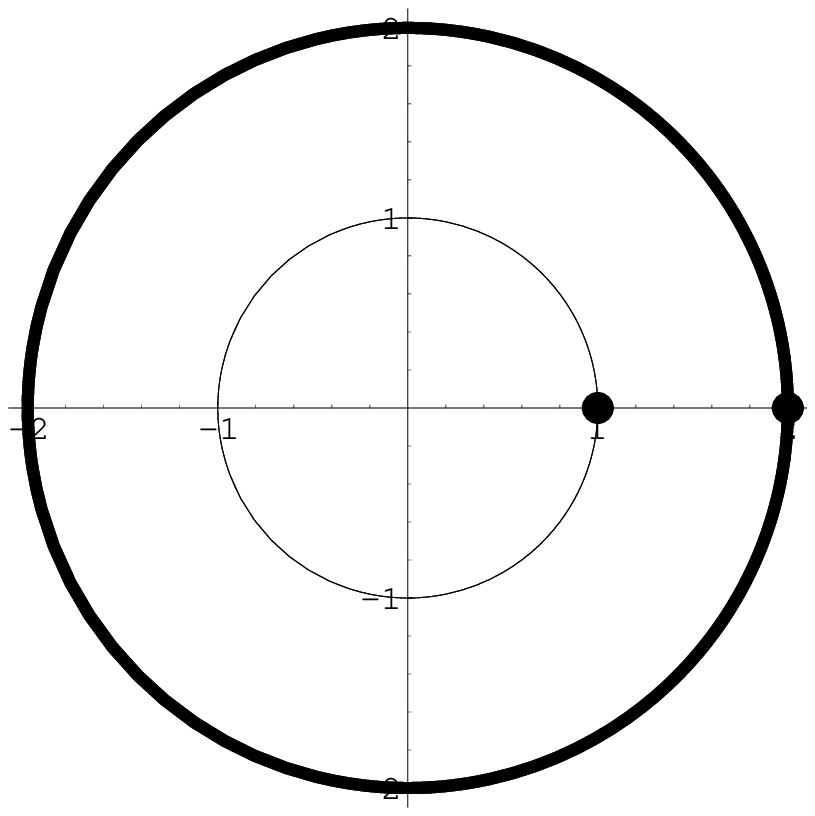}\hspace{1cm}
\includegraphics[height=3.cm]{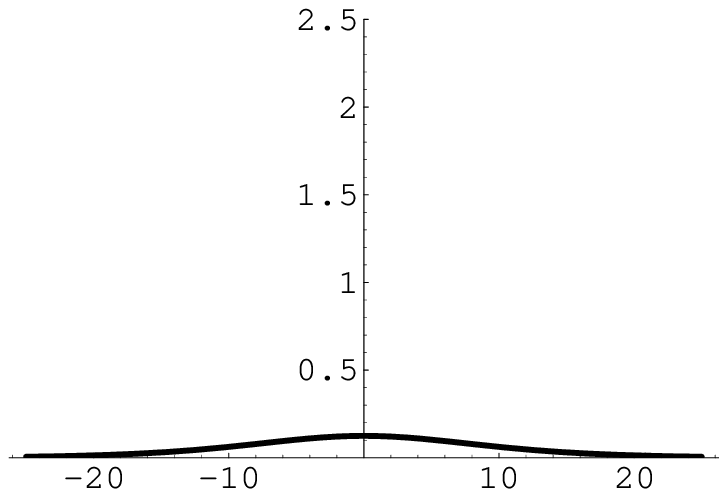}}
\caption{\small \textit{Solitary waves ${\rm K}_1^{BB}$: a) Form
Factor, b) Orbit, and c) Energy Density.}}
\end{figure}

\vspace{0.1cm}

$\bullet$ ${\rm K}_1^{AB}$: In this model there also exist kink
orbits on the separatrix curve $r_1\equiv \phi_2=0 \, \cup
\phi_1>0$. These kink orbits connect the points $A$ and $B$.
Plugging the abscissa half-axis as a trial orbit into the
first-order equations (\ref{eq:edobogo}), the following solitary
waves are found:
\[
\vec{\phi}^{{\rm K}_1^{AB}}(x)= 2\cos \left( \frac{2}{3} \arctan
e^{6\,\sqrt{2} \,\bar{x}} \right) \,\vec{e}_1 \qquad .
\]
These kinks correspond to a third kind of basic lump or particle
in the B2 model. Their energy density
\[
{\cal E}^{AB}(x)=32\, {\rm sech}^2\, 6\sqrt{2} \bar{x} \, \sin^2
\left(\frac{2}{3} \arctan e^{6\sqrt{2} \bar{x}}\right) \hspace{1cm}
, \hspace{1cm} E[{\rm
K}_1^{AB}]=\left|W^\pm(E)-W^\pm(B)\right|=\frac{22\sqrt{2}}{15}
\]
is concentrated around a point in a different fashion to the
energy densities of the iso-energetic $K_1^{AA}$ and $K_1^{BB}$
kinks, see Figure 16. $E[K_1^{AB}$ is an absolute minimum of the
energy in the ${\cal C}^{AB}$ topological sector and these kinks
are stable.

\begin{figure}[htb]
\centerline{\includegraphics[height=3.cm]{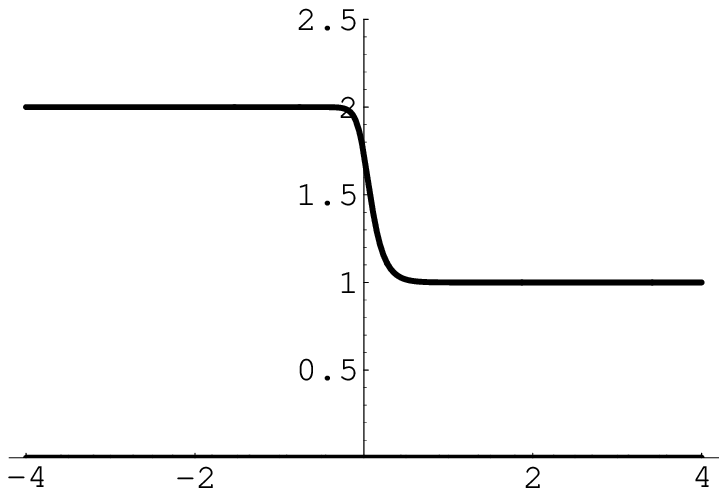}\hspace{1cm}
\includegraphics[height=3.cm]{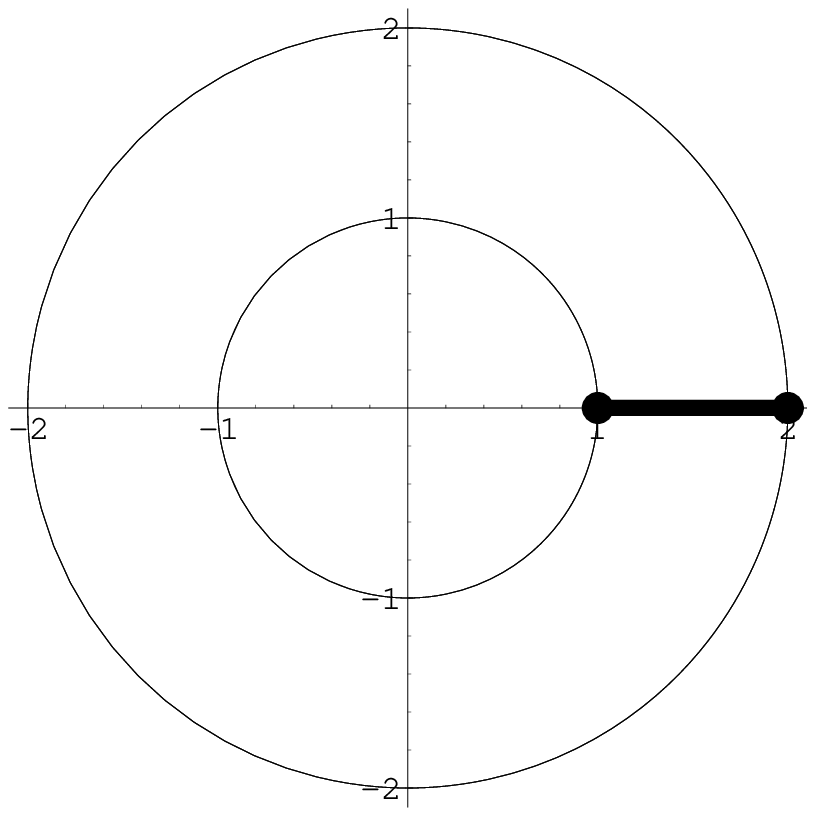}\hspace{1cm}
\includegraphics[height=3.cm]{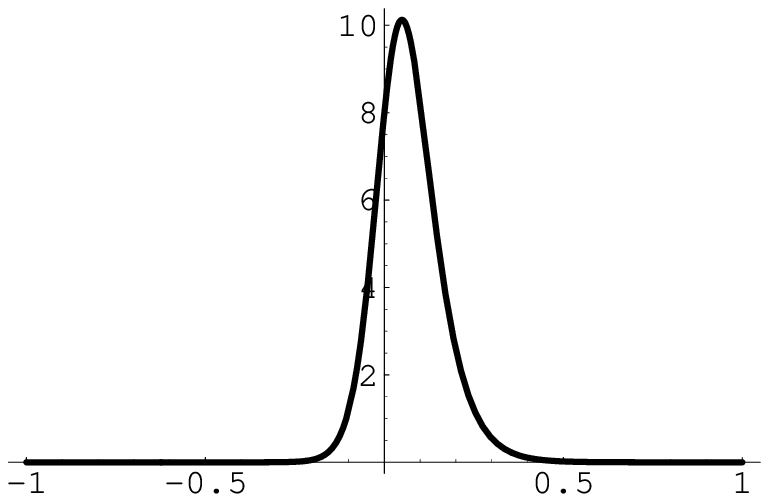}}
\caption{\small \textit{Solitary waves ${\rm K}_1^{AB}$: a) Form
Factor, b) Orbit, and c) Energy Density.}}
\end{figure}

\vspace{0.1cm}
\subsection{Solitary waves in the bulk of the moduli space}

$\bullet$ ${\rm K}_2^{AB}(\gamma_1)$: Finally, by applying the
Hamilton-Jacobi procedure we find the generic solitary wave
solutions in the bulk of the kink moduli space. There are
solutions confined inside the unit circle $C_1$ that go to the
origin. They are identical to the infinite energy solutions of the
A2 model and must be discarded as solitary waves.

There are, however, genuine solitary wave solutions confined
inside the annulus delimited by the $C_1$ and $C_2$ circles. These
kink orbits connect the two points in ${\cal M}$, $A$ and $B$. In
this case ,the Hamilton-Jacobi orbits (\ref{eq:orbgen}) and time
schedules (\ref{eq:espgen}) are provided  by the quadratures:
\begin{eqnarray*}
(-1)^\alpha \int \frac{dR}{R^2(1-R^2)(4-R^2)}-(-1)^\beta \int
\frac{d \varphi}{\sigma \sin \frac{\varphi}{2}}&=&\sqrt{2} \gamma_1 \\
\int \frac{dR}{(1-R^2)(4-R^2)}&=&\sqrt{2} \bar{x} \qquad .
\end{eqnarray*}
Integration of these equations provides the following analytic
expressions for kink orbits and form factors:
\begin{eqnarray*}
(-1)^\alpha \left[\frac{-1}{4R}+\frac{1}{6} \log
\frac{|R-2|^\frac{1}{8} (R+1)}{(R+2)^\frac{1}{8} |R-1|} \right] -
\frac{ 2 (-1)^\beta}{\sigma} \log \tan \frac{\varphi}{4}&=&\sqrt{2} \gamma_1 \\
\frac{1}{12} \log \frac{|R-2| (R+1)^2}{(R+2)(R-1)^2} &=& \sqrt{2}
\bar{x}
\end{eqnarray*}
if $1<R<2$. If $0<R<1$, the orbits above pass through the origin
and have infinite energy.

In Cartesian coordinates, these solitary wave solutions are:
\[
\vec{\phi}^{{\rm K}_2^{AB}(\gamma_1)}(x)= \phi_1^{{\rm
K}_2^{AB}(\gamma_1)}(x)\, \vec{e}_1+\phi_2^{{\rm
K}_2^{AB}(\gamma_1)}(x)\, \vec{e}_2
\]
\begin{eqnarray*}
\phi_1^{{\rm K}_2^{AB}(\gamma_1)}(x)&=& \frac{1}{\Omega_2(\bar{x})}
\left[ 1+\frac{8\, e^{2\sqrt{2} \sigma \gamma_1} \,
e^{\frac{\sigma}{2} \Omega_2(\bar{x})}}{\left( e^{\sqrt{2}\sigma
\gamma_1}\,
e^{\frac{\sigma}{4}\Omega_2(\bar{x})}+\Omega_1^\frac{\sigma}{6}(\bar{x})\,
\Lambda^\frac{\sigma}{24}(\bar{x}) \right)^2}
-\frac{8}{1+e^{-\sqrt{2} \sigma
\gamma_1}\,e^{-\frac{\sigma}{4}\Omega_2(\bar{x})}\,
\Omega_1^\frac{\sigma}{6}(\bar{x})\,
\Lambda^\frac{\sigma}{24}(\bar{x})} \right]
 \\
\phi_2^{{\rm K}_2^{AB}(\gamma_1)}(x)&=& \frac{4 \, e^\frac{\sigma
\gamma_1}{\sqrt{2}}\, e^{\frac{\sigma}{8}\Omega_2(\bar{x})}\,
\Omega_1^\frac{\sigma}{12}(\bar{x})\,
\Lambda^\frac{\sigma}{48}(\bar{x})\left[e^{\sqrt{2} \sigma
\gamma_1}\, e^{\frac{\sigma}{4}\Omega_2(\bar{x})}
-\Omega_1^\frac{\sigma}{6}(\bar{x}) \,
\Lambda^\frac{\sigma}{24}(\bar{x})  \right]}{\left[ e^{\sqrt{2}
\sigma \gamma_1} \, e^{\frac{\sigma}{4}
\Omega_2(\bar{x})}+\Omega_1^\frac{\sigma}{6}(\bar{x})\,
\Lambda^\frac{\sigma}{24}(\bar{x}) \right]^2} \qquad ,
\end{eqnarray*}
where we have made use of the notation defined in
(\ref{eq:notafun}), see also Figure 17.

\begin{figure}[htb]
\centerline{\includegraphics[height=3.cm]{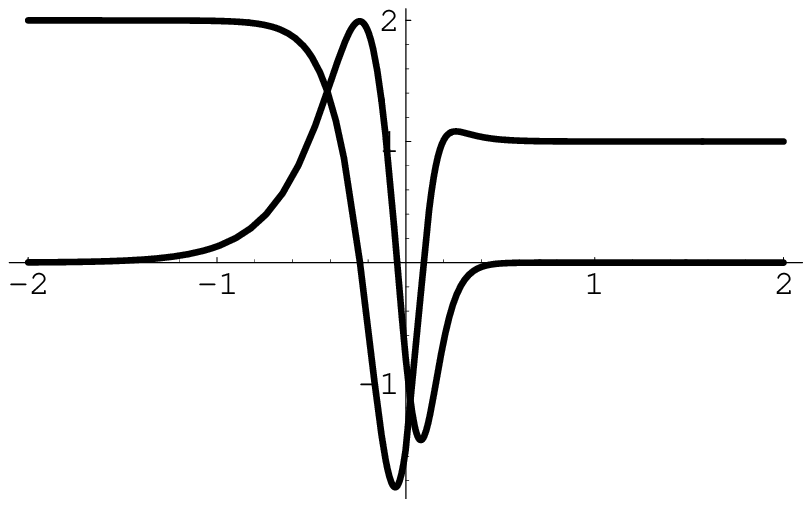}\hspace{1cm}
\includegraphics[height=3.cm]{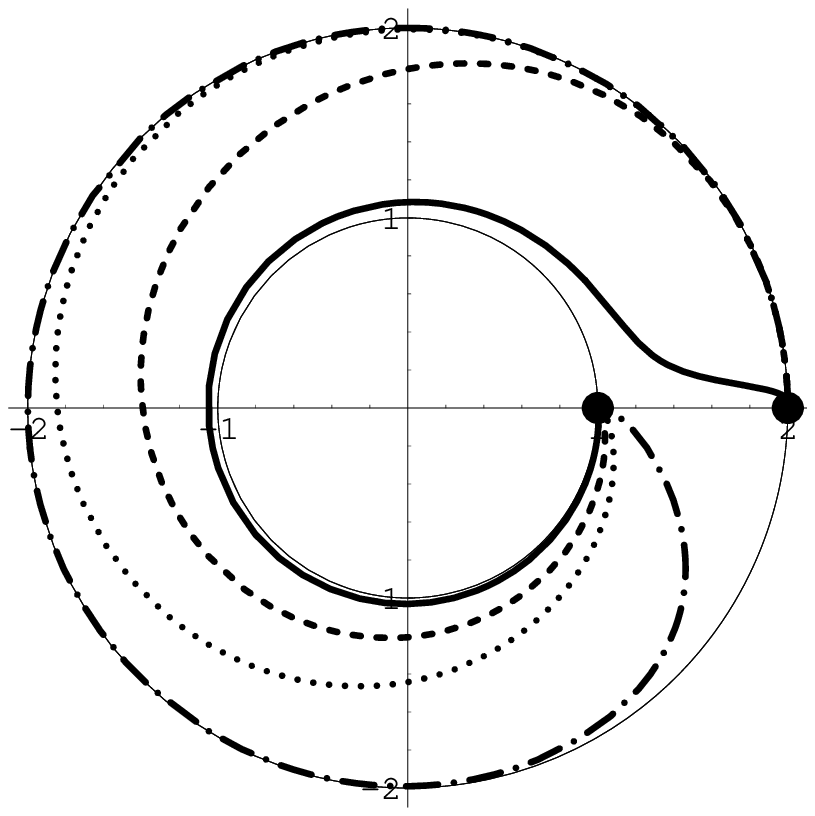}}
\caption{\small \textit{Solitary wave family ${\rm
K}_2^{AA}(\gamma_1)$: a) Form factor, and b) Orbits for several
values of $\gamma_1$.}}
\end{figure}

The distribution of the energy density reveals that solitary waves
of this type are composite. For sufficiently positive values of
the orbit parameter $\gamma_1$ we find a combination of one
$K1_1^{AB}$ and one $K_1^{AA}$ lump. This structure turns into a
combination of one $K1_1^{BB}$ and one $K_1^{AB}$ lumps for
sufficiently negative values of $\gamma_1$, see Figure 18. For
intermediate values of $\gamma_1$ the two basic kinks are
entangled. In any case, for every member of this family the
following kink energy rule holds:
\[
E[K_2^{AB}(\gamma_1)]=E[K_1^{AA}]+E[K_1^{AB}]=E[K_1^{BB}]+E[K_1^{AB}]=4\sqrt{2}\sigma+\frac{22}{15}\sqrt{2}
\qquad .
\]

\begin{figure}[htb]
\centerline{\includegraphics[height=2.3cm]{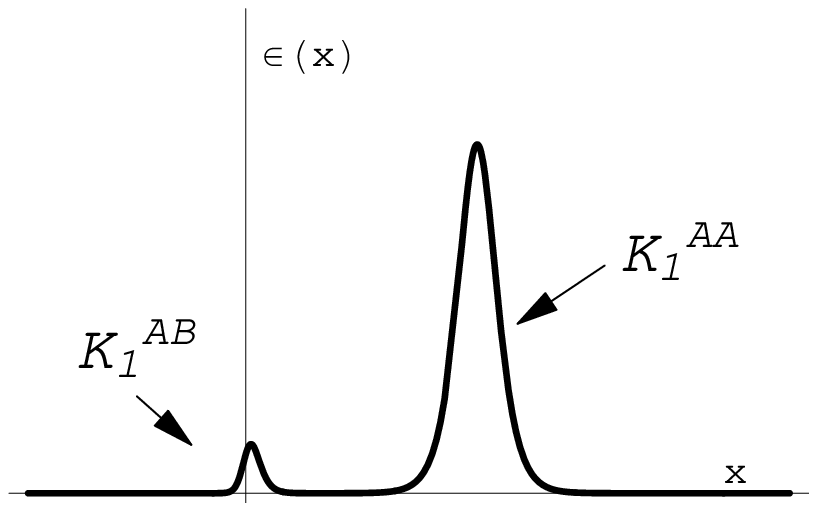}
\includegraphics[height=2.3cm]{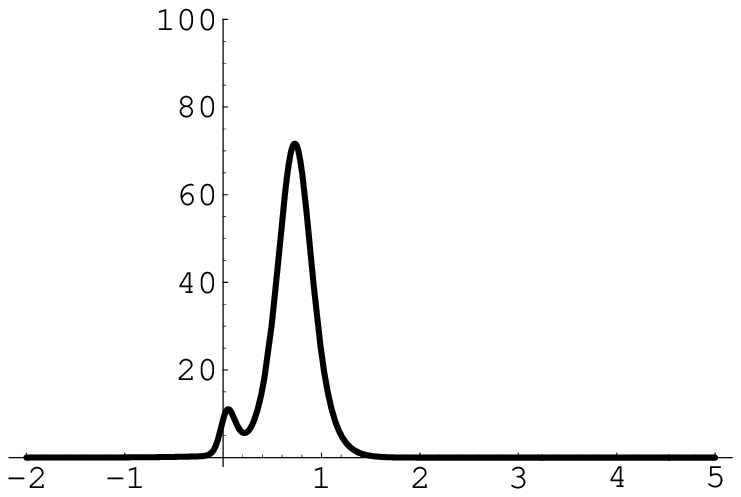}
\includegraphics[height=2.3cm]{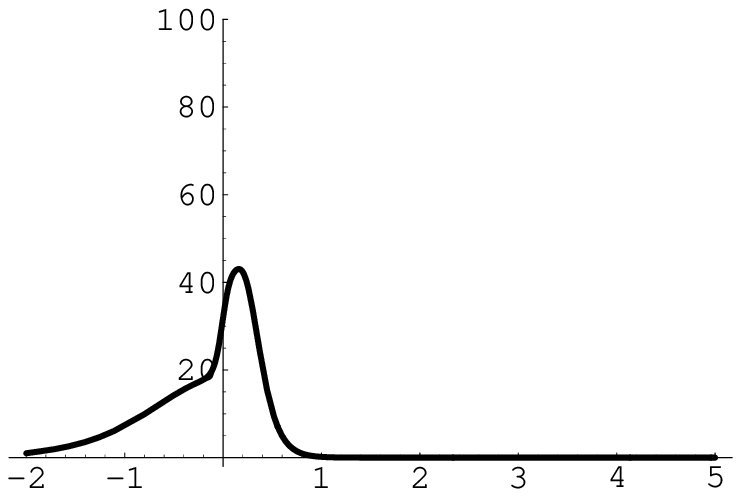}
\includegraphics[height=2.3cm]{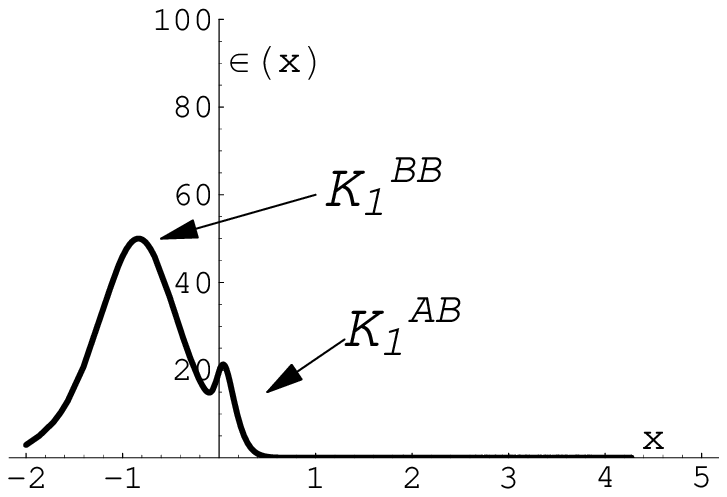}}
\caption{\small \textit{Energy density of $K_2^{AB}(\gamma_1)$
solitary waves for decreasing values of $\gamma_1$.}}
\end{figure}


\begin{thebibliography}{99}

\addcontentsline{toc}{section}{References}

\bibitem{Eschen} A.H. Eschenfelder, {\it Magnetic Bubble Technology}, (1981) Berlin, Springer-Verlag.

\bibitem{Jona} F. Jona and G. Shirane, {\it Ferroelectric Crystals}, (1993) New York, Dover; E.K. Salje, {\it Phase Transitions in Ferroelastic and Co-Elastic Crystals}, Cambridge, UK, Cambridge University Press; B.A. Strukov and A. Levanyuk, {\it Ferroelectric Phenomena in Crystals}, Berlin, Springer-Verlag.

\bibitem{Harris} J.M. Harris, {\it Poly(ethylene glycol) chemistry: Biotechnical and Biomedical Applications}, (1992) New York, Plenum.

\bibitem{Vilenkin} A. Vilenkin and E.P.S. Shellard, {\it Cosmic Strings
and Other Topological defects}, (1994) Cambridge, UK, Cambridge University
 Press.

\bibitem{Koba} S. Kobayashi, K. Koyama and J. Soda, Phys. Rev. {\bf D65}
(2002) 064014; C. Csaki, J. Erlich, T.J. Hollowood and Y. Shirman,
Nucl. Phys. {\bf B581} (2000) 309-338; M. Cvetic, Int. J. Mod. Phys.
{\bf A16} (2001) 891-899; O. DeWolfe, D.Z. Freedman, S.S. Gubser,
A.Karch, Phys. Rev. {\bf D62} (2000) 046008; H.M. Johng, H.S. Shin
and K.S. Soh, Phys. Rev. {\bf D53} (1996) 801;  N.D. Antunes, E.J.
Copeland, M. Hindmarsh and A. Lukas, {\it \lq\lq Kinky brane
worlds"}, Phys.Rev. D68 (2003) 066005; D. Bazeia, C.B. Gomes, L.
Losano, R. Menezes, Phys. Lett. B{\bf 633} (2006) 415.

\bibitem{Olive} D. Olive and E. Witten, Phys. Lett. {\bf B78} (1978) 97;
N. Seiberg and E.Witten, Nucl. Phys. {\bf B246} (1994) 19; G. Dvali
and M. Shifman, Nucl. Phys. {\bf B504} (1997) 127; G. Gibbons and P.
Townsend, Phys. Rev. Lett. {\bf 83} (1999) 172; K. Skenderis and
P.K. Townsend, Phys. Lett. {\bf B468} (1999) 46-51.

\bibitem{Drazin} P.G. Drazin, R.S. Johnson; {\it Solitons: an introduction.}, Cambridge University Press. 1989.

\bibitem{Rajaraman} R. Rajaraman, {\it Solitons and instantons. An introduction to solitons and instantons in quantum field theory}, North-Holland
Publishing Co. 1987.

\bibitem{Lohe} M.A. Lohe, Phys. Rev. D{\bf 20} (1979) 3120; L.J. Boya. J.
Casahorr\'an, Ann. Phys. {\bf 266} (1998) 63.

\bibitem{BGLM} D. Bazeia, M.A. Gonzalez Leon, L. Losano, and, J.
Mateos Guilarte, {\it Deformed Defects for Scalar Fields with
Polynomial Interactions}, Phys. Rev. {\bf D73} (2006) 105008

\bibitem{Co} S. Coleman, {\it \lq\lq There are no Goldstone bosons
in two dimensions"}, Com. Math. Phys. {\bf 31} (1973) 259-264.

\bibitem{Monto1} C. Montonen, {\it \lq\lq On solitons with an
Abelian charge in scalar field theories: (I) Classical theory and
Bohr-Sommerfeld quantization"}, Nucl. Phys. B {\bf 112} (1976)
349-357.

\bibitem{Trulli1} Sarker, S., Trullinger, S.E. y Bishop, A.R., {\it \lq\lq Solitay-wave solution for a complex one-dimensional field"}, Phys. Lett. {\bf 59A} (1976) 255-258

\bibitem{Raja1} Rajaraman, R. y Weinberg, E.J., Phys. Rev. {\bf D11} (1975) 2950

\bibitem{Trulli2} Subbaswamy, K.R. y Trullinger, S.E., {\it \lq\lq Intriguing properties of kinks in a simple model with a two-component field"}, Physica2D (1981) no. 2, 379-388.

\bibitem{Trulli3} Subbaswamy, K.R. y Trullinger, S.E., {\it \lq\lq Inestability of non topological solitons of coupled scalar field theories in two dimensions"}, Phys. Rev. {\bf D22} (1980) 1495-1496.

\bibitem{Ito1} Ito, H., {\it \lq\lq Kink energy sum rule in a two-component scalar field model of 1+1 dimensions"}, Phys. Lett. {\bf 112A} (1985) 119-123.

\bibitem{Ito2} H. Ito and H. Tasaki, {\it \lq\lq Stability theory for nonlinear Klein-Gordon Kinks and Morse's index theorem"},
Phys. Lett. A {\bf 113} (1985) 179-182.

\bibitem{J1} Mateos Guilarte, J., {\it \lq\lq Stationary phase approximation and quantum soliton families"}, Annals of Physics, {\bf 188} no. 2 (1988) 307-346.

\bibitem{Aai2} Alonso Izquierdo, A., Gonz\'alez Le\'on, M.A. y Mateos Guilarte, J., {\it \lq\lq Kink ma\-ni\-folds in (1+1)-dimensional scalar field theory"}, J. Phys. A: Math. Gen. {\bf 31} (1998), 209-229.

\bibitem{Mar1} A. Alonso Izquierdo, M.A. Gonzalez Leon, J. Mateos
Guilarte and M. de la Torre Mayado, {\it \lq\lq Kink variety in
systems of two coupled scalar fields in two space-time dimensions"},
Phys. Rev. D {\bf 65} (2002) 085012.

\bibitem{Mar} A. Alonso Izquierdo, M.A. Gonzalez Leon, J. Mateos
Guilarte and M. de la Torre Mayado, {\it \lq\lq Adiabatic motion of
two-component BPS kinks"}, Phys. Rev. D {\bf 66} (2002) 105022.

\bibitem{Modeloa} A. Alonso Izquierdo, M.A. Gonzalez Leon, M. de la
Torre Mayado, J. Mateos Guilarte, {\it \lq\lq Changing shapes:
adiabatic dynamics of composite solitary waves"}, Physica D {\bf
200} (2005) 220-241.

\bibitem{Aai3} A. Alonso Izquierdo, M. A. Gonzalez Leon, and J. Mateos Guilarte,
{\it \lq\lq Kink from dynamical systems: domain walls in a deformed
$O(N)$ linear sigma model"}, Nonlinearity {\bf 13} (2000),
1137-1169.

\bibitem{Aai4} A. Alonso Izquierdo, M. A. Gonzalez Leon, and J. Mateos Guilarte,
{\it \lq\lq Stability of kink defects in a deformed $O(3)$ linear
sigma model"}, Nonlinearity {\bf 15} (2002), 1097-1125.

\bibitem{Aai5} A. Alonso Izquierdo, J.C. Bueno S\'anchez, M. A. Gonzalez Leon, and M. de la Torre Mayado,
{\it \lq\lq Kink manifolds in a three-component scalar field
theory"}, J. Phys. A: Math. Gen. {\bf 37} (2004), 3607-3626.

\bibitem{Pere} A. Perelomov, {\sl \lq\lq Integrable Systems of Classical
Mechanics and Lie Algebras"}, Birkh\"auser, Boston MA., 1990.

\bibitem{B3} D. Bazeia, M. J. Dos Santos and R. F. Ribeiro,
 {\it \lq\lq Solitons in systems of coupled scalar fields"},
Phys. Lett. A {\bf 208} (1995) 84-88.

\bibitem{B5} D. Bazeia, J. R. S. Nascimento, R. F. Ribeiro and D. Toledo,
{\it \lq\lq Soliton stability in systems of two real scalar
fields"}, J. Phys. A {\bf 30} (1997) 8157-8166.

\bibitem{Bogo} E. B. Bogomol'nyi,
{\it \lq\lq The stability of classical solutions"}, Sov. J. Nucl.
Phys. {\bf 24} (1976) 449-454.

\end{thebibliography}
\end{document}